\documentclass[draft]{agujournal2019}
\usepackage{url}
\usepackage{lineno}
\usepackage[inline]{trackchanges}
\usepackage{soul}
\usepackage{apacite}
\usepackage{amssymb}
\usepackage{gensymb}
\usepackage{mathrsfs,amsmath}
\usepackage{pifont}

\usepackage{multirow}
\usepackage{graphicx}

\usepackage[utf8x]{inputenc}
\usepackage{array}
\usepackage{diagbox}

\usepackage{tabularx}
\usepackage{xcolor,colortbl}
\usepackage{multirow, caption, booktabs}

\usepackage{hhline}
\usepackage{colortbl}
\usepackage{rotating}

\begin{document}

\title{ Nonlinear Interactions of Planetary-Scale Waves in Mesospheric Winds Observed at 52°N Latitude and Two Longitudes}

\authors{Maosheng He\affil{1,2},
Jeffrey M. Forbes\affil{3},
Gunter Stober \affil{4},
Christoph Jacobi\affil{5},
Guozhu Li\affil{6,7},
Libo Liu\affil{6,8},
Jiyao Xu\affil{1}
}

\affiliation{1}{Key Laboratory of Solar Activity and Space Weather, National Space Science Center, Chinese Academy of Sciences, Beijing, China}
\affiliation{2}{Hainan National Field Science Observation and Research Observatory for Space Weather, National Space Science Center, Chinese Academy of Sciences, Beijing, China}
\affiliation{3}{Ann \& H. J. Smead Department of Aerospace Engineering Sciences, University of Colorado, Boulder, USA}
\affiliation{4}{Institute of Applied Physics \& Oeschger Center for Climate Change Research, Microwave Physics, University of Bern, Bern, Switzerland}
\affiliation{5}{Institute for Meteorology, Leipzig University, Leipzig, Germany}
\affiliation{6}{Beijing national observatory of space environment, Institute of Geology and Geophysics, Chinese Academy of Sciences, Beijing, China}
\affiliation{7}{College of Earth and Planetary Sciences, University of Chinese Academy of Sciences, Beijing, China}
\affiliation{8}{Heilongjiang Mohe National Observatory of Geophysics, Beijing, China}

\correspondingauthor{Maosheng He}{hemaosheng@nssc.ac.cn}

\begin{keypoints}
\item Planetary wave normal modes drive multi-day oscillations, showing April-October seasonality and statistical SSW associations (not one-to-one)
\item  First evidence of frequency and zonal wavenumber matching for over 10 secondary waves of nonlinear interactions among planetary-scale waves
\item Non-migrating components dominate the winter 24-h tide and summer 8-h tide, attributed to the nonlinear interactions
\end{keypoints}
\begin{abstract}
Nine years of mesospheric wind data from two meteor radars at 52°N latitude were analyzed to investigate planetary waves (PWs) and tides by estimating their zonal wavenumber through longitudinal phase differences. Our results reveal that PW normal modes (NMs) primarily drive multi-day oscillations, showing seasonal variability and statistical associations with Sudden Stratospheric Warming (SSW) events. Specifically, a significant 6-day NM emerges in April, followed by predominant 4- and 2-day NMs until June, with peaks of 2-, 4-, and 6-day NMs spanning July to October. Furthermore, our study provides the first observational verification of frequency and zonal wavenumber of over ten secondary waves from nonlinear interactions among planetary-scale waves. One notable finding is the prevalence of non-migrating components in winter 24-hour and summer 8-hour tides, attributed to these nonlinear interactions. Our findings underscore the diverse nonlinear dynamics of planetary-scale waves, triggering a variety of periodic oscillations.
\end{abstract}

\section*{Plain Language Summary}

We analyzed nine years of mid-latitude middle atmospheric wind data at two longitudes. By examining longitudinal phase differences, we identified planetary-scale waves and their sources. Most multi-day oscillations are linked to planetary wave normal modes, influenced by atmospheric mechanical properties. Certain normal modes were prominent from April to October, while others correlated with winter’s sudden stratospheric warming events. These modes, along with tides and stationary planetary waves, interact nonlinearly, generating secondary waves of non-normal mode multi-day oscillations and sun-asynchronous tides. Our two-station method reveals for the first time secondary waves of stationary planetary waves, which share the same frequencies as their parent waves, making them indistinguishable in single-station spectral analyses.  This phenomenon might explain the prevalence of sun-asynchronous components in the 24-hour winter and 8-hour summer tides, which are exceptions to the typically dominant sun-synchronous tides. Additionally, we report for the first time the interactions of normal modes with non-migrating and 8-hour tides. Our findings highlight the extensive and varied nonlinear behaviors of planetary-scale waves, resulting in a broad spectrum of oscillations.

\clearpage

\section{Introduction}
Extratropical planetary waves (PWs), also referred to as Rossby waves, play a significant role in the dynamics of rotating fluids, where the meridional gradient of the Coriolis force acts as the restoring force \cite{Zaqarashvili2021,Holton2013}. In  Earth's atmosphere,  PWs exhibit exceptional behavior in the form of normal modes (NMs), intricately linked to the resonant properties of the medium. Constrained predominantly within westerly winds due to  Earth's rotation, PWs manifest a westward phase velocity relative to their carrying fluids, a consequence of the conservation of absolute vorticity. These  principles are illustrated in Figure 1 of \citeA{He2022NC}. The 	temporal spectrum of PW NMs spans discretely  scales from a few days to several weeks, with periods of 2, 4, 5--6, 10, 16, and 25--28 days.

Ground-based observations have proven pivotal in delineating the seasonality of PWs \cite<e.g.,>[]{hocking2011a, Jacobi2008,Jiang2008, sridharan2006, gong2018a, yu2019a}. However, most of these studies relied on single-station analyses,  lacking the capacity to measure horizontal scales. Space-based observations, on the other hand, have contributed to investigating the climatology of PWs with constraints on zonal wavenumber \cite<e.g.,>[ respectively]{Gu2013, yamazaki2021, Forbes2017a, Forbes2015, McDonald2011} for waves with periods of 2, 4, 6, 10, and 16 days. Nevertheless, these observations, mostly collected in quasi-sun-synchronous orbits and using single-satellite analyses, and are thus susceptible to ambiguities from aliasing, as discussed by \citeA{Forbes2017a}. Distinguishing NMs from secondary waves resulting from interactions between NMs and migrating tides is challenging, as elaborated in Section 1 of \citeA{He2021g}.

Recently, using mesospheric wind data observed from two meteor radars at 52°N latitude, \citeA{He2022NC} demonstrated that the dominant 2--32-day fluctuations during the 2019 New Year sudden stratospheric warming event (SSW) can be attributed to either PW NMs or their nonlinear behaviors. This analysis determines frequency and zonal wavenumber directly from observations, unlike satellite-based climatological interpretations that often fit individual waves using predetermined frequencies and wavenumbers. Here, we extend this event study to a 9-year window,  elucidating PWs' seasonal variations   and  statistical occurrences during SSWs. The association between NMs and SSWs remains debated. While some studies affirm this association \cite{Pancheva2008jgr, Yamazaki2019}, while others challenge it \cite{Sassi2012}.
In additional, our zonal wavenumber estimations facilitate  identifying the secondary waves of stationary PW (sPW) interactions with migrating tides and NMs which are indistinguishable in single-station spectral analyses as they have the same frequencies as the parent tides and NMs.

\section{Data Analysis} \label{sec:data}

The data utilized in this study are comprise of observations of mesosphere winds spanning from 2012 to 2020, obtained from two meteor radars  situated at comparable latitudes but seperated in longitude by 109$^\circ$. One station is positioned in Mohe at coordinates 122$^\circ$E, 53.5$^\circ$N \cite{Yu2013}, while the second is situated in Collm, coordinates 13.0$^\circ$E, 51.3$^\circ$N \cite{Jacobi2012,Stober2021, Stober_2022_3DVAR+DIV}. Employing the methodology delineated by \cite{Hocking2001,Holdsworth:2004} for the SKiYMET and ATRAD meteor radars, respectively, zonal and meridional wind components ($u$ and $v$) are computed on an hourly basis within an altitude range spanning from 80.5 km to 95.5 km.

\subsection{Estimation of Zonal Wavenumber } \label{sec:PDT}

Following the methodology employed in Figure 2 of \citeA{He2022NC}, the cross spectra between Mohe and Collm are computed for horizontal winds, as illustrated in Figure \ref{fig:Fig1}. This procedure involves computing cross-wavelet spectra of the zonal ($u$) and meridional ($v$) wind components separately at each altitude level, followed by summing these spectra and subsequently averaging the resultant sum across all altitudes. The cross spectrum enables the estimation of the zonal wavenumber ($s$) associated with the underlying wave, employing the phase difference technique (PDT).

Similar to other discrete spectral analyses, the PDT involves several assumptions, including adherence to the Nyquist sampling theorem. This principle dictates that the sampling frequency in both temporal and spatial domains must exceed twice the maximum frequency and wavenumber inherent in the signal. However, validating these assumptions through models is intrinsically insufficient, as sufficient validation necessitates complete knowledge of the signal, particularly its components at frequencies and wavenumbers surpassing the Nyquist rates, which often pose significant modeling challenges. In the PDT framework, prior knowledge allows a partial relaxation of the assumption regarding the Nyquist rate in space, as discussed in the context of the "long-wave assumption" by \citeA{He2020GRLb}. Notably, this assumption's relaxation varies across scenarios, often guided by a predefined set of potential $s$. The current study uses the relaxations defined in \citeA{He2022NC} and \citeA{He2024} in dealing with the inter- and intra-diurnal spectra, respectively. These relaxations primarily aim to identify the simplest explanation among potential aliasing candidates beyond the Nyquist rate.

Another crucial assumption concerns the constraints of linear systems, where solving for a given number of unknowns necessitates an equal number of equations. Sampling from $N$ locations enables the estimation of wave amplitudes for up to $N$ different predefined wavenumbers within a specified time-frequency grid. In existing literature, the prevailing approach to studying the upper mesosphere involves $N$ = 1 analyses, conducted via single-station or -satellite spectral methods. These analyses operate under the single-wave assumption, presuming the presence of only one predominant wave within each discernible time-frequency grid. Consequently, in $N$ = 1 analyses, estimating wavenumber in the detector's intrinsic coordinate system is infeasible, and wavenumber explanation is based on prior knowledge. Observations from an additional location allow for verifying the applicability of the single-wave assumption, for example, by comparing the phase difference between the locations with its expectation associated with the expected wavenumber. In practical applications in the PDT, this verification does not rely on the wavenumber expectation. Since different wave modes typically exhibit distinct evolutions in the time and frequency domains, the superposition of different wave modes is  accompanied by phase differences that vary with time and frequency. This characteristic can be used to identify breaches in the single-wave assumption, as illustrated in Section \ref{sec:NIs}. Conversely, a stable phase difference over time and frequency indicates that the single-wave assumption is being met. This phase difference is a function of the wavenumber and the separation between locations, thereby allowing for wavenumber estimation. Readers are referred to \citeA{He2023} for a review of the PDT approach implemented in a series of works involving both case and statistical studies, including PDT cross-validations using independently paired stations and consistent comparisons with estimations from three- or four-station analyses \cite{He2020, He2021gb}.

\subsection{Composite Analysis} \label{sec:Composite}

The spectra illustrated in Figure \ref{fig:Fig1} are composited with regard to the day of  year and to the central days of SSWs are delineated in Figures \ref{fig:Fig2}a--\ref{fig:Fig2}d and \ref{fig:Fig2}e--\ref{fig:Fig2}h, respectively. The identified central days, marked by vertical dashed lines in Figure \ref{fig:Fig1}, are as follows: 17-01-2012, 11-01-2013, 05-01-2015, 06-03-2016, 02-02-2017, 14-02-2018, and 31-12-2018. These reference days are determined based on the occurrences of polar vortex weakening events (PVWs), discerned through the analysis of eastward wind variation at 48 km  altitude and  70$^\circ$N latitude, utilizing MERRA reanalysis data  \cite{Zhang2014a}.

\section{Results} \label{sec:Results}

Figure \ref{fig:Fig1}a presents the cross-wavelet spectrum for periods $T=40$h--$36$d from 2012 to 2020,   in which the color intensity denotes the amplitudes, while  hue indicates the phase difference between the Mohe and Collm radars, aiding in the estimation of the zonal wavenumber $s$ for dominant waves. Red, green, and blue hues correspond to wavenumbers $s=1$, 2, and 3, respectively.  The estimated $s$, labeled at some dominant peaks as Arabic numerals "1" and "2" for $s=1$ and 2,  shows dependence on the period $T$, with most red peaks occurring at periods $T>5$d, while green peaks predominate at $3<T<5$d. For $T<3$d, spectral peaks vary in color and predominantly appear in summer.

\subsection{Seasonal Composite of Multi-day Spectra}

Seasonal variations are detailed in the composite analyses in Figure \ref{fig:Fig2}a, where the most intense spectral peaks emerge in the summer half-year (from April to October), mostly at periods $T\leq6$d. The summer spectral peaks manifest a distinct V-shaped distribution. From March to June, the predominant peaks exhibit monotonically decreasing periods (from $T=6$ days sequentially decreasing to 4, then 2 days), while the wavenumber sequentially increases from $s=1$ to 2, then 3. Subsequently, from July to October, the predominant peaks show an opposite trend: the period monotonically increases while the wavenumber  decreases. This period-wavenumber elucidation of PW seasonality, covering periods greater than 3 days, is reported here for the first time.

Figure \ref{fig:Fig2}a furthermore unveils a magenta peak in July--August, at a period $T<3$d. This peak may involve multiple quasi-two-day waves, as discussed in \cite{He2021g}, to which readers are referred for a comprehensive understanding of year-to-year and seasonal variability in the spectral segment with $T<3$d. We will not reiterate these aspects here.

\subsection{SSW Composite of Multi-day Spectra}

Unlike summer, when the spectrum exhibits significant seasonality, the winter half-year (from November to February) manifests sporadic spectral peaks with  diminished amplitudes. Notably, a conspicuous pattern is not readily discernible in Figure \ref{fig:Fig2}a during the winter half-year. Instead, the winter spectrum prominently reveals a dependence on SSWs, as evidenced, for instance, in Figure \ref{fig:Fig1}a for the early months of 2013, early 2018, and late 2018, where the spectra exhibit pronounced multi-peaks. The SSW composite spectrum manifests peaks at $T=16$, 10, 8, 6, 4, and 2 days, as indicated by the digits or symbols '1', '1', '2', '1', '?' and '?' in Figure \ref{fig:Fig2}e, respectively. The first four digits identify the wavenumber $s$ diagnosed through the PDT, while the question symbols "?" denote failures of wavenumber diagnosis. These failures are attributed to the mixing of waves with different wavenumbers, a discussion of which is provided in Section 4.

\subsection{Near-24-, 12-, and 8-hour Spectra and Their Composites}

Figures \ref{fig:Fig1}b--\ref{fig:Fig1}d depict spectra similar to Figure \ref{fig:Fig1}a but spanning the frequency ranges  $1.00\pm0.25$ cycles per day (cpd), $2.00\pm0.25$ cpd, and $3.00\pm0.25$ cpd, respectively. The composites of these spectra are illustrated in Figures \ref{fig:Fig2}b--\ref{fig:Fig2}d and \ref{fig:Fig2}f--\ref{fig:Fig2}h, with respect to the calendar month and the SSW reference days, respectively. In these representations and for most years, the 24-, 12-, and 8-hour spectra predominantly appear in red, green, and blue, indicating modes $s=1$, $2$, and $3$, respectively. These modes correspond to the diurnal, semidiurnal, and terdiurnal migrating tides.\\
Diverging from these migrating tidal signatures and as indicated by the string '2e' in Figure \ref{fig:Fig2}b, the 24-hour spectrum exhibits a prevalence of green hues in November--December, suggesting $s=2$, which is a signature of a non-migrating tide. Similarly, a non-migrating signature appears also at the period $T=8$h. As indicated by the question symbol "?" in Figure \ref{fig:Fig2}d, the 8-hour peak is not in blue colors in May--August but rather exhibits a complex color attributed in Section 4 to mixing of waves with different wavenumbers.\\
In Figure \ref{fig:Fig2}g, following the vertical dashed line at $T=12.4$h, a green peak emerges, indicating $s=2$, and implying a migrating semidiurnal lunar tide \cite{Forbes2012,He2019} , though alternative explanations were also argued \cite{Caspel_2023_NAVGEM_tides,fuller-rowell2016}.\\
Additionally, flanking the 24-, 12-, and 8-hour migrating tidal signatures in Figures \ref{fig:Fig1} and \ref{fig:Fig2} are scattered off-24-, 12-, and 8-hour spectral peaks, which will be further discussed in the subsequent section.

\section{Discussions} \label{sec:Discussion}

Section \ref{sec:Results} illustrated temporal variations of oscillations at the periods of multiple days, and near-24, -12, and -8 hours. The spectra are composted to highlight the seasonal variations and the statistical associations with SSW, and PDT is used to diagnose the zonal wavenumbers of underlying waves. The current section discusses the associated dynamics. Hereafter, we employ a pair of numbers enclosed in square brackets to denote waves, $[\frac{f}{1 \text{cpd}}, s]$, where $\frac{f}{1 \text{cpd}}$  identifies the frequency in cpd, and $s$ signifies the zonal wavenumber.

\subsection{ Normal Modes} \label{sec:NMs}
\begin{sidewaysfigure}
\centering
\includegraphics[width=23.5cm]{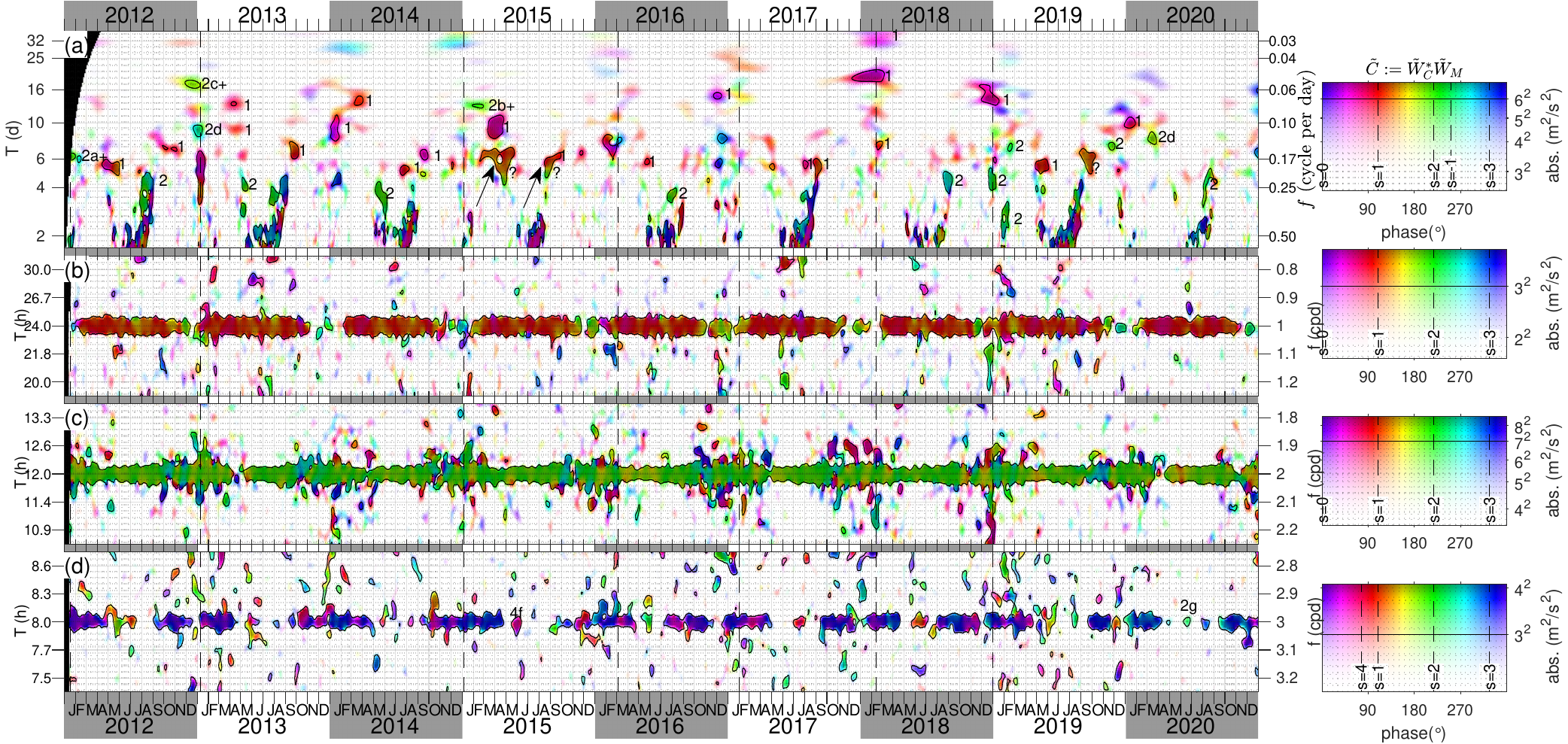}
\caption{
Altitude-averaged (80.5--95.5km) cross-wavelet spectra	between Mohe and Collm across four frequency bands:  (a) \(f\) = 0.03-0.50 cpd, (b) 1.00$\pm$0.25 cpd, (c) 2.00$\pm$0.25 cpd, and (d)  3.00$\pm$0.25 cpd.  The spectral color density indicates the magnitude of the spectrum, while the hue reflects the phase difference between our Mohe and Collm stations.
The color density of the spectra denotes the spectral magnitude while the colour denotes the phase difference between our Mohe and Collm stations, which is a function of the zonal wavenumber $s$ of the underlying wave and longitudinal separation between our Mohe and Collm stations.
Arabic numerals (1, 2, 3, etc.) denote nearby spectral peaks, representing the estimated zonal wavenumber \(s\) of associated underlying waves. Letters 'a+', 'b+', etc., following some Arabic numerals indicate labeled spectral peaks resulting from nonlinear interactions specified in Lines (a+), (b+), etc., in Table  \ref{tab:title}. 	In each panel, the solid black and white lines represent two reference isolines, either specified by the horizontal line in the corresponding colormap or by the white number on the isolines when the isoline is saturated in the colormap.}
\label{fig:Fig1}
\end{sidewaysfigure}

\begin{figure}
\centering
\includegraphics[width=15.5cm]{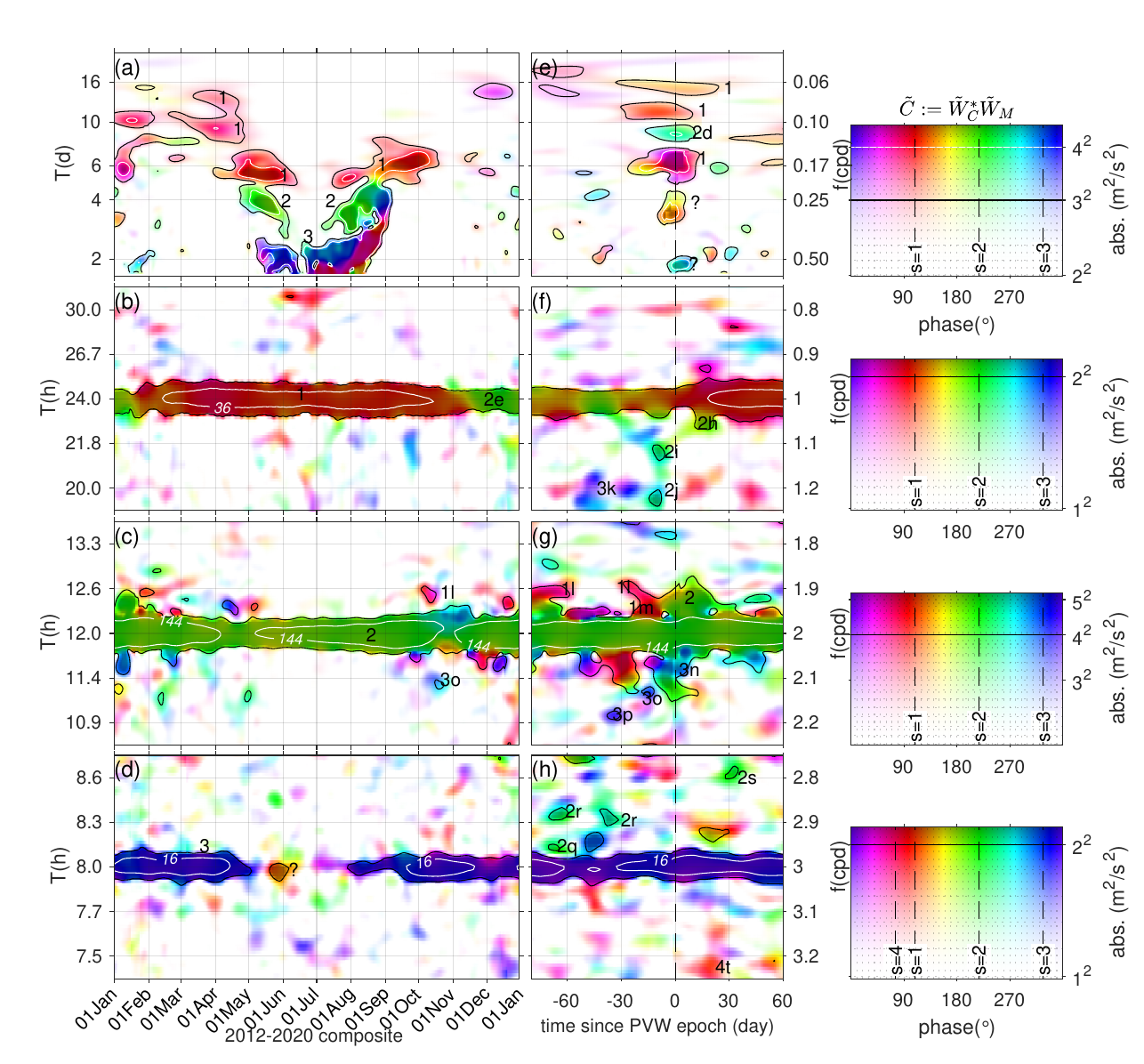}
\caption{Composite analysis of spectra displayed in Figure \ref{fig:Fig1}a with respect to (a) day of year and (e) the central day of SSWs.  Panels (b,f), (c,g), and (d,h) show similar plots to (a,e) but for the near-24h, -12h, and -8h spectra displayed in Figures  \ref{fig:Fig1}b,  \ref{fig:Fig1}c and  \ref{fig:Fig1}d, respectively. Arabic numerals (1, 2, 3, etc.) annotate  spectral peaks, representing the estimated zonal wavenumber \(s\) of associated underlying waves. Letters 'd', 'e', etc., following the    numerals denote  annotated  spectral peaks resulting from nonlinear interactions as detailed in Lines (d), (e), etc.,  in Table  \ref{tab:title}. The question symbols "?" indicate spectral  peaks that are attributable to superpositions of waves with multiple zonal wavenumbers See section 4 for details.  In each panel, the solid black and white lines represent two reference isolines, either specified by the horizontal line in the corresponding colormap or by the white number on the isolines when the isoline is saturated in the colormap.}
\label{fig:Fig2}
\end{figure}

Most of the multi-day spectral peaks mentioned above can be attributed to the 16-, 10-, 6-, 4-, and 2-day PW  NMs or Rossby gravity NMs, $ [\frac{1}{16},1], [\frac{1}{10},1], [\frac{1}{6},1], [\frac{1}{4},2]$, and $[\frac{24}{50},3]$ \cite<for details, see>[respectively]{Forbes1995a,Forbes1995b,Forbes1995b,yamazaki2021,He2021g}. Note that these frequencies represent the NMs in average observations, which differs from the intrinsic frequencies anticipated by classical theory due to the Doppler shift associated with the background wind.
Hence, observations of these waves are often prefaced with "quasi-", such as quasi-16-day and quasi-6-day waves. For instance, the $[\frac{1}{16},1]$ mode can be anticipated in the range $11<T<20$ days based on \citeA{Salby1981}, providing an explanation for all spectral peaks with wavenumber $s=1$ within this period range in Figure  \ref{fig:Fig1}a. The  enhancements of the quasi-6-, 4-, and 2-day NMs in the summer-half year have been investigated individually in various datasets \cite<e.g.,>[respectively]{Talaat2001, yamazaki2021, He2021g}.  Here, Figure  \ref{fig:Fig2}a reveals  the seasonal variation of NMs with a single dataset in a single analysis. Apart from these NMs, Figure \ref{fig:Fig1}a also displays signatures of other NMs. For example, the red spectral peak ($s=1$) near $T=30$d in February 2018 and the green spectral peak ($s=2$) near $T=7$d in February 2019 can be attributed to the 28-day \cite{Zhao2019} and 7-day NMs \cite{Pogoreltsev2002a}, respectively.\\

In Figures \ref{fig:Fig1}a and \ref{fig:Fig2}a, a distinct winter-summer  contrast emerges in the frequency distribution of NMs. Short-period NMs ($T <= 6$ days) predominantly manifest during  summer, while longer-period NMs ($T > 6$ days) are more prevalent in winter. This seasonal contrast can be qualitatively explained  in terms of  the longitudinal average of the zonal wind ($\overline{u}$) and its influence on the PW phase velocity ($c$)  as encapsulated by the inequalities $0<\overline{u}-c<\overline{u}_{\text{cri}}$   \cite{charney1961}.
The first inequality $0<\overline{u}-c$ dictates westward propagation of PWs  with respect to the background flow, determined by Earth's rotation  and the associated  north-south gradient of relative vorticity \cite{Rossby1939,andrews1987,He2022NC}.  The second inequality, defining a threshold (the critical velocity $\overline{u}_{\text{cri}}$) of $\overline{u}-c$, ensures a reversible distortion of potential vorticity contours. Beyond this threshold, the distortion becomes pronounced and irreversible, impeding PW propagation and leading to breaking.  In the middle atmosphere at middle latitudes, the difference in $\overline{u}$ between winter and summer  primarily manifests as $  \overline{u}_\text{summer}< \overline{u}_\text{winter} $,  with westerly $\overline{u}_\text{winter}  > 0$ during winter and easterly $\overline{u}_\text{summer}  < 0$ during summer \cite{jacobi2009b}.  Combining both $0<\overline{u}-c<\overline{u}_{\text{cri}}$ and $\overline{u}_\text{summer}  < \overline{u}_\text{winter} $, it is reasonable to anticipate $\overline{c}_\text{summer}<\overline{c}_\text{winter}$, implying that the winter atmosphere allows NMs with larger  $c$ compared to those permitted in summer. Note that since $\overline{c}$ is predominantly negative, larger $\overline{c}$ during winter than summer indicates that waves in winter travel at a slower absolute velocity than in summer.		In Figures \ref{fig:Fig1}a and \ref{fig:Fig2}a, the NMs prevalent in the winter half-year include, for instance, the 28-, 16-, and 10-day NMs associated with $c =$ --9.7, --17.0, and --27.2 m/s, respectively, which are larger (namely slower) than the $c =$ --45.4, --34.0, and --45.4 m/s of the 6-, 4-, and 2-day NMs predominantly observed in the summer half-year. It is noteworthy that the phase velocity is defined in the Earth-fixed frame as $c := \frac{2\pi a \cos \phi}{-sT}$, where $a = 6370$ km defines the Earth's radius, $\phi$ represents the latitude, and the negative symbol preceding $s$ indicates that a positive $c$ or $u$ denotes eastward motion, opposite to the positive $s$ direction. Also note that for the qualitative discussion above, we  disregarded the dependence of $\overline{u}_{\text{cri}}$ on NMs' horizontal wavelength.

During SSWs, $\overline{u}$ undergoes a rapid transition from westerly to easterly within a short span of time. This reversal isn't uniform; $\overline{u}$ may stall at certain values for a while, resembling potentially winter- or summer-like conditions. Moreover, the progression of $\overline{u}$ reversal varies across events, facilitating potentially the occurrence of  winter- or summer-preferred NMs, as evidenced in Figure \ref{fig:Fig2}e.
While sPWs are known to play a critical role in initiating SSWs \cite{Matsuno1971},  the involvement of NMs in SSWs remains contentious. Some studies support the link between NMs and SSWs \cite{Pancheva2008jgr,Yamazaki2019,He2020jd}, while others do not 	\cite{Sassi2012}. The results presented in our Figures \ref{fig:Fig1} and \ref{fig:Fig2} demonstrate a clear  association from a statistical perspective rather than a one-to-one correspondence. Figure \ref{fig:Fig1}a shows that not every SSW is associated with NMs, and not all winter PWs are accompanied by SSWs.
Our study contributes to resolving the debate over the association between  NMs and SSWs.

While NMs effectively account for the majority of significant spectral peaks observed in Figures \ref{fig:Fig2}a and \ref{fig:Fig2}e,  a limited number of exceptions occur. The first two exceptions are the near-2-day spectral peaks, in magenta  and cyan in Figures \ref{fig:Fig2}a  and \ref{fig:Fig2}e, respectively, both of which are  denoted  by the question marks "?" in the Figures.  Each of these peaks may comprise multiple wave components. Specifically, the magenta peak may consist of waves with wavenumber $s=4$ and --2, while the cyan peak may involve $s=3$ and 2 \cite<for details see>[respectively]{He2021g,He2021gb}. A third exception is discerned in the peak at $T=4$ days in Figure \ref{fig:Fig2}e, which may also entangle multiple spectral signatures, e.g., the red and green spectral peaks around SSWs 2012/2013 and 2018/2019 in Figure \ref{fig:Fig1}a, respectively. The green color suggests $s=2$ and is attributable to the NM $[\frac{1}{4},2]$, whereas the red color might result from the contamination of the NM $[\frac{1}{6},1]$. This speculation of contamination is founded on Figure  \ref{fig:Fig1}a, where, during the 2012/2013 SSW, a singular spectral peak of almost uniform color spans from $T=4$ day to $T=6$ day, encompassing phase signals of both 6- and 4-day waves in principle. The last exception is the green peak at $T=8$ days in Figure \ref{fig:Fig2}e which will be detailed in Section \ref{sec:NIs}.

\subsection{ Nonlinear interactions} \label{sec:NIs}

\subsubsection*{Non-NM multi-day spectral peaks}

In addition to the previously mentioned NMs, Figure  \ref{fig:Fig1}a contains spectral peaks with $s=2$ that cannot be explained solely in terms of NMs but rather through secondary nonlinear interactions between different PWs. For an in-depth understanding of secondary non-linearity, readers are referred to \citeA{He2022NC}. Through such interactions, two waves may generate two new waves, denoted as:
$[\frac{f_1}{1 \text{cpd}}, s_1] \pm [\frac{f_2}{1 \text{cpd}}, s_2] = [\frac{f_1 \pm f_2}{1 \text{cpd}}, s_1 \pm s_2]$.
Here, '1' and '2' index the parent waves, while '+' and '--' denote  secondary waves, known as upper sideband (USB) and lower sideband (LSB), respectively. In Figure  \ref{fig:Fig1}a, the green peaks indicated by '2a+', '2b+', and'2c+'  are associated with estimation $s=2$  and could be explained as the USB of   interactions between NMs $[\frac{1}{6},1]$, $[\frac{1}{10},1]$, and $[\frac{1}{16},1]$with sPWs $[0,1]$ as specified in Lines (a+), (b+), and (c+) in Table \ref{tab:title}, respectively.
Here, a potential aliasing for the $s=2$ estimation is $s=-1$, differing by only 33$^\circ$ in the color phase on the color map in Figure \ref{fig:Fig1}a.
These  $s=-1$ aliases should not be neglected as they could be explained as the LSB of the nonlinear interactions of the previous NMs with the sPW  as specified in Lines (a--), (b--), and (c--) in Table \ref{tab:title}.  Both [0,1] and [0,2] are frequently observed, though the [0,1] amplitude is statistically significantly stronger than the [0,2] \cite{shi2021,smith1997}.  Superposition between the green LSB- or SBU-like SWs ($s=-1$ or 2) with the red NM peaks ($s=1$) could further explain the spectral peaks arrowed in Figure \ref{fig:Fig1}a in 2015. These peaks exhibit complex colors, a combination of red, green, and yellow, which could be attributed to the overlapping of a red and a green peak with distinct evolution in the time and frequency domains.  Such superposition between the SW-like wave and the NMs was reported in satellite data \cite{Ma2024}.

In addition, the green peak labeled as '2d' at $T=8$ days in Figure \ref{fig:Fig2}e could be attributed to secondary harmonic generation of an NM as reported by \citeA{He2022NC} and specified in  Line (d) in Table \ref{tab:title}. Figure \ref{fig:Fig1}a might also incorporate signatures of more nonlinear interactions, and our intention is to discuss the most typical ones rather than to provide an exhaustive list.

\subsubsection*{Non-migrating tidal signatures}
Section 3.3 illustrated that while the 24-, 12-, and 8-hour spectra are dominated by migrating components [1,1], [2,2], and [3,3], there are also signatures of non-migrating tides.
Specifically, the 24-hour green peak in December in Figure \ref{fig:Fig2}b can be explained in term of the interaction between the diurnal migrating tide and a sPW, [0,1], as specified in Line (e) Table \ref{tab:title}. Similarly, the 8-hour green and red peaks in June 2020 and May 2015, indicated by the strings '2g' and '4f' in Figure 1d, can be explained as the LSB and USB of the interactions between the tide [3,3]  and the sPW [0,1], as specified in Lines (f, g) in Table \ref{tab:title}, respectively. The superposition of these secondary waves could account for the non-migrating tidal signature in Figure \ref{fig:Fig2}d that is indicated by the question symbol (?).

\subsubsection*{Tidal sidebands}

In Figures \ref{fig:Fig2}b--\ref{fig:Fig2}d, sporadic isolated spectral peaks are observed at off-12-hour periods. These peaks are commonly interpreted as secondary waves of interactions between tides and NMs. In Figure \ref{fig:Fig2}g, the red peaks, labeled '1m' and '1l' and associated with \(s=1\), signify the LSBs of interactions between the 12-hour migrating tide and NMs  as specified in Lines (m,l) of Table \ref{tab:title}. In contrast, the blue peaks, labeled '3n', '3o', and '3p' and associated with \(s=3\), represent the corresponding USBs specified in Lines (n--p) of Table \ref{tab:title}. Similarly, the green peaks labeled '2h', '2i', and '2j' in Figure \ref{fig:Fig2}f, and the green and red peaks labeled '2q', '2r', '2s', and '4t' in Figure \ref{fig:Fig2}h, signify the nonlinear sidebands of diurnal and terdiurnal tides, as  specified in Lines (h--k) and (q--t)  of Table \ref{tab:title}, respectively. Additionally, these '3k' and '4t'  signatures  in Figures \ref{fig:Fig2}f and \ref{fig:Fig2}h are attributable to non-migrating tidal sidebands of the interactions specified in Lines (f) and (h) of Table \ref{tab:title}, respectively.

In existing literature using ground observations, the existence of tidal sidebands is primarily inferred from frequency matching. Here, we demonstrate the existence of sidebands based on both frequency and zonal wavenumber matching. Compared to the more commonly reported off-12-hour tidal sidebands, off-24-hour and off-8-hour sidebands are rarely, if ever, observed in the midlatitude mesosphere. Specifically, we present observational evidence for the NM nonlinear interactions with non-migrating tides and 8-hour tides for the first time. The novelty of these findings is highlighted in Table \ref{tab:title}.

\section{Conclusion}

Utilizing meteor radar observations spanning nine years at two  longitudes and 52°N latitude, this study investigates planetary-scale waves in mesospheric winds. By examining zonal wavenumbers $s$ across various time scales (multi-day, near-24-hour, -12-hour, and -8-hour), we distinguished NMs from other PWs, identified migrating and non-migrating tides, and diagnosed a variety of novel nonlinear interactions. Our statistical analysis revealed that PWs were primarily linked to NMs, which exhibit a specific period/wavenumber seasonality during the summer half-year and a winter association with SSWs. Notably, April showcased a prominent 6-day NM (with $s=1$), followed by a prevalence of 4- and 2-day NMs (with $s=2$ and $s=3$, respectively) extending through June. Subsequent peaks in 2-, 4-, and 6-day NMs (with $s=3$, $s=2$, and $s=1$, respectively) were observed from July to October. Our insights into seasonal variations are derived from observational determinations of frequency and zonal wavenumber, contrasting with satellite observations that often rely on fitting individual waves using predetermined frequencies and wavenumbers. Our findings on the statistical association between NMs and SSWs significantly contribute to resolving the ongoing debate on this topic. Additionally, for the first time, we identified evidence of frequency and zonal wavenumber matching for more than ten secondary waves resulting from nonlinear interactions among NMs (16-, 10-, and 6-day), tides (diurnal, semidiurnal, and terdiurnal; migrating and non-migrating), and sPWs These interactions represent three novel categories involving parent waves of sPWs, terdiurnal tides, and non-migrating tides. The sPW interactions could explain our finding that non-migrating tidal amplitudes surpass the corresponding migrating tides observed in the winter diurnal tide and summer terdiurnal tide. These non-migrating signatures are notable exceptions, as migrating components generally dominate diurnal, semidiurnal, and terdiurnal tides for most months.

\section*{Open Research}
The hourly wind data from Mohe is provided by the Data Center for Geophysics, National Earth System Science Data Sharing Infrastructure at Beijing National Observatory of Space Environment, Institute of Geology and Geophysics, Chinese Academy of Sciences). The hourly wind data at Mohe and Collm  are available in the World Data Center \cite{He2020Data1} for Geophysics, Beijing,  and Harvard Dataverse  \cite{He2023Data} .

\acknowledgments
This work is supported by the Chinese Meridian Project and  the Specialized Research Fund for State Key Laboratory in China. Christoph Jacobi is supported by  Deutsche Forschungsgemeinschaft grants JA 863/47-1.  Gunter Stober is a member of the Oeschger Center for Climate Change Research.  The authors declare that they have no competing financial interests.

\bibliography{library}

\begin{thebibliography}{}

\bibitem [\protect \citeauthoryear {%
Andrews%
, Holton%
\BCBL {}\ \BBA {} Leovy%
}{%
Andrews%
\ \protect \BOthers {.}}{%
{\protect \APACyear {1987}}%
}]{%
andrews1987}
\APACinsertmetastar {%
andrews1987}%
\begin{APACrefauthors}%
Andrews, D\BPBI G.%
, Holton, J\BPBI R.%
\BCBL {}\ \BBA {} Leovy, C\BPBI B.%
\end{APACrefauthors}%
\unskip\
\newblock
\APACrefYear{1987}.
\newblock
\APACrefbtitle {Middle {Atmosphere} {Dynamics}} {Middle {Atmosphere}
  {Dynamics}}.
\newblock
\APACaddressPublisher{}{Academic Press}.
\newblock
\APACrefnote{Google-Books-ID: N1oNurYZefAC}
\PrintBackRefs{\CurrentBib}

\bibitem [\protect \citeauthoryear {%
Charney%
\ \BBA {} Drazin%
}{%
Charney%
\ \BBA {} Drazin%
}{%
{\protect \APACyear {1961}}%
}]{%
charney1961}
\APACinsertmetastar {%
charney1961}%
\begin{APACrefauthors}%
Charney, J\BPBI G.%
\BCBT {}\ \BBA {} Drazin, P\BPBI G.%
\end{APACrefauthors}%
\unskip\
\newblock
\APACrefYearMonthDay{1961}{}{}.
\newblock
{\BBOQ}\APACrefatitle {Propagation of planetary-scale disturbances from the
  lower into the upper atmosphere} {Propagation of planetary-scale disturbances
  from the lower into the upper atmosphere}.{\BBCQ}
\newblock
\APACjournalVolNumPages{Journal of Geophysical Research
  (1896-1977)}{66}{1}{83-109}.
\newblock
\begin{APACrefURL}
  \url{https://agupubs.onlinelibrary.wiley.com/doi/abs/10.1029/JZ066i001p00083}
  \end{APACrefURL}
\newblock
\begin{APACrefDOI} \doi{https://doi.org/10.1029/JZ066i001p00083}
  \end{APACrefDOI}
\PrintBackRefs{\CurrentBib}

\bibitem [\protect \citeauthoryear {%
Forbes%
}{%
Forbes%
}{%
{\protect \APACyear {1995}}%
}]{%
Forbes1995b}
\APACinsertmetastar {%
Forbes1995b}%
\begin{APACrefauthors}%
Forbes, J\BPBI M.%
\end{APACrefauthors}%
\unskip\
\newblock
\APACrefYearMonthDay{1995}{}{}.
\newblock
{\BBOQ}\APACrefatitle {{Tidal and planetary waves}} {{Tidal and planetary
  waves}}.{\BBCQ}
\newblock
\APACjournalVolNumPages{Geophysical Monograph Series}{87}{}{67--87}.
\newblock
\begin{APACrefURL}
  \url{http://www.agu.org/books/gm/v087/GM087p0067/GM087p0067.shtml}
  \end{APACrefURL}
\newblock
\begin{APACrefDOI} \doi{10.1029/GM087p0067} \end{APACrefDOI}
\PrintBackRefs{\CurrentBib}

\bibitem [\protect \citeauthoryear {%
Forbes%
\ \protect \BOthers {.}}{%
Forbes%
\ \protect \BOthers {.}}{%
{\protect \APACyear {1995}}%
}]{%
Forbes1995a}
\APACinsertmetastar {%
Forbes1995a}%
\begin{APACrefauthors}%
Forbes, J\BPBI M.%
, Hagan, M\BPBI E.%
, Miyahara, S.%
, Vial, F.%
, Manson, A\BPBI H.%
, Meek, C\BPBI E.%
\BCBL {}\ \BBA {} Portnyagin, Y\BPBI I.%
\end{APACrefauthors}%
\unskip\
\newblock
\APACrefYearMonthDay{1995}{}{}.
\newblock
{\BBOQ}\APACrefatitle {{Quasi 16-day oscillation in the mesosphere and lower
  thermosphere}} {{Quasi 16-day oscillation in the mesosphere and lower
  thermosphere}}.{\BBCQ}
\newblock
\APACjournalVolNumPages{Journal of Geophysical Research}{100}{D5}{9149}.
\newblock
\begin{APACrefURL} \url{http://doi.wiley.com/10.1029/94JD02157}
  \end{APACrefURL}
\newblock
\begin{APACrefDOI} \doi{10.1029/94JD02157} \end{APACrefDOI}
\PrintBackRefs{\CurrentBib}

\bibitem [\protect \citeauthoryear {%
Forbes%
\ \BBA {} Zhang%
}{%
Forbes%
\ \BBA {} Zhang%
}{%
{\protect \APACyear {2012}}%
}]{%
Forbes2012}
\APACinsertmetastar {%
Forbes2012}%
\begin{APACrefauthors}%
Forbes, J\BPBI M.%
\BCBT {}\ \BBA {} Zhang, X.%
\end{APACrefauthors}%
\unskip\
\newblock
\APACrefYearMonthDay{2012}{}{}.
\newblock
{\BBOQ}\APACrefatitle {{Lunar tide amplification during the January 2009
  stratosphere warming event: Observations and theory}} {{Lunar tide
  amplification during the January 2009 stratosphere warming event:
  Observations and theory}}.{\BBCQ}
\newblock
\APACjournalVolNumPages{Journal of Geophysical Research: Space
  Physics}{117}{12}{1--13}.
\newblock
\begin{APACrefDOI} \doi{10.1029/2012JA017963} \end{APACrefDOI}
\PrintBackRefs{\CurrentBib}

\bibitem [\protect \citeauthoryear {%
Forbes%
\ \BBA {} Zhang%
}{%
Forbes%
\ \BBA {} Zhang%
}{%
{\protect \APACyear {2015}}%
}]{%
Forbes2015}
\APACinsertmetastar {%
Forbes2015}%
\begin{APACrefauthors}%
Forbes, J\BPBI M.%
\BCBT {}\ \BBA {} Zhang, X.%
\end{APACrefauthors}%
\unskip\
\newblock
\APACrefYearMonthDay{2015}{}{}.
\newblock
{\BBOQ}\APACrefatitle {{Quasi-10-day wave in the atmosphere}} {{Quasi-10-day
  wave in the atmosphere}}.{\BBCQ}
\newblock
\APACjournalVolNumPages{Journal of Geophysical
  Research}{120}{21}{11,079--11,089}.
\newblock
\begin{APACrefDOI} \doi{10.1002/2015JD023327} \end{APACrefDOI}
\PrintBackRefs{\CurrentBib}

\bibitem [\protect \citeauthoryear {%
Forbes%
\ \BBA {} Zhang%
}{%
Forbes%
\ \BBA {} Zhang%
}{%
{\protect \APACyear {2017}}%
}]{%
Forbes2017a}
\APACinsertmetastar {%
Forbes2017a}%
\begin{APACrefauthors}%
Forbes, J\BPBI M.%
\BCBT {}\ \BBA {} Zhang, X.%
\end{APACrefauthors}%
\unskip\
\newblock
\APACrefYearMonthDay{2017}{}{}.
\newblock
{\BBOQ}\APACrefatitle {{The quasi-6 day wave and its interactions with solar
  tides}} {{The quasi-6 day wave and its interactions with solar
  tides}}.{\BBCQ}
\newblock
\APACjournalVolNumPages{Journal of Geophysical Research: Space
  Physics}{122}{4}{4764--4776}.
\newblock
\begin{APACrefURL} \url{http://doi.wiley.com/10.1002/2017JA023954}
  \end{APACrefURL}
\newblock
\begin{APACrefDOI} \doi{10.1002/2017JA023954} \end{APACrefDOI}
\PrintBackRefs{\CurrentBib}

\bibitem [\protect \citeauthoryear {%
Fuller-Rowell%
\ \protect \BOthers {.}}{%
Fuller-Rowell%
\ \protect \BOthers {.}}{%
{\protect \APACyear {2016}}%
}]{%
fuller-rowell2016}
\APACinsertmetastar {%
fuller-rowell2016}%
\begin{APACrefauthors}%
Fuller-Rowell, T\BPBI J.%
, Fang, T\BHBI W.%
, Wang, H.%
, Matthias, V.%
, Hoffmann, P.%
, Hocke, K.%
\BCBL {}\ \BBA {} Studer, S.%
\end{APACrefauthors}%
\unskip\
\newblock
\APACrefYearMonthDay{2016}{}{}.
\newblock
{\BBOQ}\APACrefatitle {Impact of {Migrating} {Tides} on {Electrodynamics}
  {During} the {January} 2009 {Sudden} {Stratospheric} {Warming}} {Impact of
  {Migrating} {Tides} on {Electrodynamics} {During} the {January} 2009 {Sudden}
  {Stratospheric} {Warming}}.{\BBCQ}
\newblock
\BIn{} \APACrefbtitle {Ionospheric {Space} {Weather}} {Ionospheric {Space}
  {Weather}}\ (\BPGS\ 163--174).
\newblock
\APACaddressPublisher{}{American Geophysical Union (AGU)}.
\newblock
\begin{APACrefURL}
  [{2024-02-04}]\url{https://onlinelibrary.wiley.com/doi/abs/10.1002/9781118929216.ch14}
  \end{APACrefURL}
\newblock
\APACrefnote{Section: 14 \_eprint:
  https://onlinelibrary.wiley.com/doi/pdf/10.1002/9781118929216.ch14}
\newblock
\begin{APACrefDOI} \doi{10.1002/9781118929216.ch14} \end{APACrefDOI}
\PrintBackRefs{\CurrentBib}

\bibitem [\protect \citeauthoryear {%
Gong%
\ \protect \BOthers {.}}{%
Gong%
\ \protect \BOthers {.}}{%
{\protect \APACyear {2018}}%
}]{%
gong2018a}
\APACinsertmetastar {%
gong2018a}%
\begin{APACrefauthors}%
Gong, Y.%
, Li, C.%
, Ma, Z.%
, Zhang, S.%
, Zhou, Q.%
, Huang, C.%
\BDBL {}Ning, B.%
\end{APACrefauthors}%
\unskip\
\newblock
\APACrefYearMonthDay{2018}{}{}.
\newblock
{\BBOQ}\APACrefatitle {Study of the {Quasi}-5-{Day} {Wave} in the {MLT}
  {Region} by a {Meteor} {Radar} {Chain}} {Study of the {Quasi}-5-{Day} {Wave}
  in the {MLT} {Region} by a {Meteor} {Radar} {Chain}}.{\BBCQ}
\newblock
\APACjournalVolNumPages{Journal of Geophysical Research:
  Atmospheres}{123}{17}{9474--9487}.
\newblock
\begin{APACrefURL}
  [{2024-02-09}]\url{https://onlinelibrary.wiley.com/doi/abs/10.1029/2018JD029355}
  \end{APACrefURL}
\newblock
\APACrefnote{\_eprint:
  https://onlinelibrary.wiley.com/doi/pdf/10.1029/2018JD029355}
\newblock
\begin{APACrefDOI} \doi{10.1029/2018JD029355} \end{APACrefDOI}
\PrintBackRefs{\CurrentBib}

\bibitem [\protect \citeauthoryear {%
Gu%
\ \protect \BOthers {.}}{%
Gu%
\ \protect \BOthers {.}}{%
{\protect \APACyear {2013}}%
}]{%
Gu2013}
\APACinsertmetastar {%
Gu2013}%
\begin{APACrefauthors}%
Gu, S\BPBI Y.%
, Li, T.%
, Dou, X.%
, Wu, Q.%
, Mlynczak, M\BPBI G.%
\BCBL {}\ \BBA {} Russell, J\BPBI M.%
\end{APACrefauthors}%
\unskip\
\newblock
\APACrefYearMonthDay{2013}{}{}.
\newblock
{\BBOQ}\APACrefatitle {{Observations of Quasi-Two-Day wave by TIMED/SABER and
  TIMED/TIDI}} {{Observations of Quasi-Two-Day wave by TIMED/SABER and
  TIMED/TIDI}}.{\BBCQ}
\newblock
\APACjournalVolNumPages{Journal of Geophysical Research
  Atmospheres}{118}{4}{1624--1639}.
\newblock
\begin{APACrefDOI} \doi{10.1002/jgrd.50191} \end{APACrefDOI}
\PrintBackRefs{\CurrentBib}

\bibitem [\protect \citeauthoryear {%
He%
}{%
He%
}{%
{\protect \APACyear {2020}}%
}]{%
He2020Data1}
\APACinsertmetastar {%
He2020Data1}%
\begin{APACrefauthors}%
He, M.%
\end{APACrefauthors}%
\unskip\
\newblock
\APACrefYearMonthDay{2020}{}{}.
\newblock
\APACrefbtitle {The hourly mesospheric winds over Mohe between 2012 and 2019.}
  {The hourly mesospheric winds over mohe between 2012 and 2019.}
\newblock
\APACaddressPublisher{}{WDC for Geophysics, Beijing}.
\newblock
\begin{APACrefDOI} \doi{10.12197/2020GA016} \end{APACrefDOI}
\PrintBackRefs{\CurrentBib}

\bibitem [\protect \citeauthoryear {%
He%
}{%
He%
}{%
{\protect \APACyear {2023}}%
}]{%
He2023}
\APACinsertmetastar {%
He2023}%
\begin{APACrefauthors}%
He, M.%
\end{APACrefauthors}%
\unskip\
\newblock
\APACrefYearMonthDay{2023}{{\APACmonth{04}}}{}.
\newblock
{\BBOQ}\APACrefatitle {Planetary-scale {MLT} waves diagnosed through
  multi-station methods: a review} {Planetary-scale {MLT} waves diagnosed
  through multi-station methods: a review}.{\BBCQ}
\newblock
\APACjournalVolNumPages{Earth, Planets and Space}{75}{1}{63}.
\newblock
\begin{APACrefURL}
  [{2023-07-11}]\url{https://doi.org/10.1186/s40623-023-01808-5}
  \end{APACrefURL}
\newblock
\begin{APACrefDOI} \doi{10.1186/s40623-023-01808-5} \end{APACrefDOI}
\PrintBackRefs{\CurrentBib}

\bibitem [\protect \citeauthoryear {%
He%
\ \BBA {} Chau%
}{%
He%
\ \BBA {} Chau%
}{%
{\protect \APACyear {2019}}%
}]{%
He2019}
\APACinsertmetastar {%
He2019}%
\begin{APACrefauthors}%
He, M.%
\BCBT {}\ \BBA {} Chau, J\BPBI L.%
\end{APACrefauthors}%
\unskip\
\newblock
\APACrefYearMonthDay{2019}{}{}.
\newblock
{\BBOQ}\APACrefatitle {{Mesospheric semidiurnal tides and near-12{\&}thinsp;h
  waves through jointly analyzing observations of five specular meteor radars
  from three longitudinal sectors at boreal midlatitudes}} {{Mesospheric
  semidiurnal tides and near-12{\&}thinsp;h waves through jointly analyzing
  observations of five specular meteor radars from three longitudinal sectors
  at boreal midlatitudes}}.{\BBCQ}
\newblock
\APACjournalVolNumPages{Atmospheric Chemistry and Physics}{19}{9}{5993--6006}.
\newblock
\begin{APACrefURL} \url{https://doi.org/10.5194/acp-19-5993-2019}
  \end{APACrefURL}
\newblock
\begin{APACrefDOI} \doi{10.5194/acp-19-5993-2019} \end{APACrefDOI}
\PrintBackRefs{\CurrentBib}

\bibitem [\protect \citeauthoryear {%
He%
, Chau%
\BCBL {}\ \protect \BOthers {.}}{%
He%
, Chau%
\BCBL {}\ \protect \BOthers {.}}{%
{\protect \APACyear {2020}}%
}]{%
He2020GRLb}
\APACinsertmetastar {%
He2020GRLb}%
\begin{APACrefauthors}%
He, M.%
, Chau, J\BPBI L.%
, Forbes, J\BPBI M.%
, Thorsen, D.%
, Li, G.%
, Siddiqui, T\BPBI A.%
\BDBL {}Hocking, W\BPBI K.%
\end{APACrefauthors}%
\unskip\
\newblock
\APACrefYearMonthDay{2020}{}{}.
\newblock
{\BBOQ}\APACrefatitle {{Quasi-10-Day Wave and Semidiurnal Tide Nonlinear
  Interactions During the Southern Hemispheric SSW 2019 Observed in the
  Northern Hemispheric Mesosphere}} {{Quasi-10-Day Wave and Semidiurnal Tide
  Nonlinear Interactions During the Southern Hemispheric SSW 2019 Observed in
  the Northern Hemispheric Mesosphere}}.{\BBCQ}
\newblock
\APACjournalVolNumPages{Geophysical Research Letters}{47}{23}{e2020GL091453}.
\newblock
\begin{APACrefURL} \url{https://doi.org/10.1029/2020GL091453} \end{APACrefURL}
\newblock
\begin{APACrefDOI} \doi{10.1029/2020GL091453} \end{APACrefDOI}
\PrintBackRefs{\CurrentBib}

\bibitem [\protect \citeauthoryear {%
He%
, Chau%
\BCBL {}\ \protect \BOthers {.}}{%
He%
, Chau%
\BCBL {}\ \protect \BOthers {.}}{%
{\protect \APACyear {2021}}%
}]{%
He2021gb}
\APACinsertmetastar {%
He2021gb}%
\begin{APACrefauthors}%
He, M.%
, Chau, J\BPBI L.%
, Forbes, J\BPBI M.%
, Zhang, X.%
, Englert, C\BPBI R.%
, Harding, B\BPBI J.%
\BDBL {}Makela, J\BPBI J.%
\end{APACrefauthors}%
\unskip\
\newblock
\APACrefYearMonthDay{2021}{}{}.
\newblock
{\BBOQ}\APACrefatitle {{Quasi-2-Day Wave in Low-Latitude Atmospheric Winds as
  Viewed From the Ground and Space During January–March, 2020}} {{Quasi-2-Day
  Wave in Low-Latitude Atmospheric Winds as Viewed From the Ground and Space
  During January–March, 2020}}.{\BBCQ}
\newblock
\APACjournalVolNumPages{Geophysical Research Letters}{48}{13}{}.
\newblock
\begin{APACrefURL} \url{https://doi.org/10.1029/2021GL093466} \end{APACrefURL}
\newblock
\begin{APACrefDOI} \doi{10.1029/2021GL093466} \end{APACrefDOI}
\PrintBackRefs{\CurrentBib}

\bibitem [\protect \citeauthoryear {%
He%
, Chau%
, Hall%
\BCBL {}\ \protect \BOthers {.}}{%
He%
, Chau%
, Hall%
\BCBL {}\ \protect \BOthers {.}}{%
{\protect \APACyear {2018}}%
}]{%
He2018g}
\APACinsertmetastar {%
He2018g}%
\begin{APACrefauthors}%
He, M.%
, Chau, J\BPBI L.%
, Hall, C\BPBI M.%
, Tsutsumi, M.%
, Meek, C.%
\BCBL {}\ \BBA {} Hoffmann, P.%
\end{APACrefauthors}%
\unskip\
\newblock
\APACrefYearMonthDay{2018}{}{}.
\newblock
{\BBOQ}\APACrefatitle {{The 16-Day Planetary Wave Triggers the SW1-Tidal-Like
  Signatures During 2009 Sudden Stratospheric Warming}} {{The 16-Day Planetary
  Wave Triggers the SW1-Tidal-Like Signatures During 2009 Sudden Stratospheric
  Warming}}.{\BBCQ}
\newblock
\APACjournalVolNumPages{Geophysical Research Letters}{45}{22}{12,631--12,638}.
\newblock
\begin{APACrefURL} \url{https://doi.org/10.1029/2018GL079798} \end{APACrefURL}
\newblock
\begin{APACrefDOI} \doi{10.1029/2018GL079798} \end{APACrefDOI}
\PrintBackRefs{\CurrentBib}

\bibitem [\protect \citeauthoryear {%
He%
, Chau%
, Stober%
\BCBL {}\ \protect \BOthers {.}}{%
He%
, Chau%
, Stober%
\BCBL {}\ \protect \BOthers {.}}{%
{\protect \APACyear {2018}}%
}]{%
He2018j}
\APACinsertmetastar {%
He2018j}%
\begin{APACrefauthors}%
He, M.%
, Chau, J\BPBI L.%
, Stober, G.%
, Li, G.%
, Ning, B.%
\BCBL {}\ \BBA {} Hoffmann, P.%
\end{APACrefauthors}%
\unskip\
\newblock
\APACrefYearMonthDay{2018}{}{}.
\newblock
{\BBOQ}\APACrefatitle {{Relations Between Semidiurnal Tidal Variants Through
  Diagnosing the Zonal Wavenumber Using a Phase Differencing Technique Based on
  Two Ground-Based Detectors}} {{Relations Between Semidiurnal Tidal Variants
  Through Diagnosing the Zonal Wavenumber Using a Phase Differencing Technique
  Based on Two Ground-Based Detectors}}.{\BBCQ}
\newblock
\APACjournalVolNumPages{Journal of Geophysical Research:
  Atmospheres}{123}{8}{4015--4026}.
\newblock
\begin{APACrefURL} \url{https://doi.org/10.1002/2018JD028400} \end{APACrefURL}
\newblock
\begin{APACrefDOI} \doi{10.1002/2018JD028400} \end{APACrefDOI}
\PrintBackRefs{\CurrentBib}

\bibitem [\protect \citeauthoryear {%
He%
\ \BBA {} Forbes%
}{%
He%
\ \BBA {} Forbes%
}{%
{\protect \APACyear {2022}}%
}]{%
He2022NC}
\APACinsertmetastar {%
He2022NC}%
\begin{APACrefauthors}%
He, M.%
\BCBT {}\ \BBA {} Forbes, J\BPBI M.%
\end{APACrefauthors}%
\unskip\
\newblock
\APACrefYearMonthDay{2022}{{\APACmonth{12}}}{}.
\newblock
{\BBOQ}\APACrefatitle {Rossby Wave Second Harmonic Generation Observed in the
  Middle Atmosphere} {Rossby wave second harmonic generation observed in the
  middle atmosphere}.{\BBCQ}
\newblock
\APACjournalVolNumPages{Nature Communications}{13}{1}{7544}.
\newblock
\begin{APACrefDOI} \doi{10.1038/s41467-022-35142-3} \end{APACrefDOI}
\PrintBackRefs{\CurrentBib}

\bibitem [\protect \citeauthoryear {%
He%
, Forbes%
\BCBL {}\ \protect \BOthers {.}}{%
He%
, Forbes%
\BCBL {}\ \protect \BOthers {.}}{%
{\protect \APACyear {2020}}%
}]{%
He2020}
\APACinsertmetastar {%
He2020}%
\begin{APACrefauthors}%
He, M.%
, Forbes, J\BPBI M.%
, Chau, J\BPBI L.%
, Li, G.%
, Wan, W.%
\BCBL {}\ \BBA {} Korotyshkin, D\BPBI V.%
\end{APACrefauthors}%
\unskip\
\newblock
\APACrefYearMonthDay{2020}{mar}{}.
\newblock
{\BBOQ}\APACrefatitle {{High-Order Solar Migrating Tides Quench at SSW Onsets}}
  {{High-Order Solar Migrating Tides Quench at SSW Onsets}}.{\BBCQ}
\newblock
\APACjournalVolNumPages{Geophysical Research Letters}{47}{6}{1--8}.
\newblock
\begin{APACrefURL} \url{https://doi.org/10.1029/2019GL086778} \end{APACrefURL}
\newblock
\begin{APACrefDOI} \doi{10.1029/2019GL086778} \end{APACrefDOI}
\PrintBackRefs{\CurrentBib}

\bibitem [\protect \citeauthoryear {%
He%
\ \protect \BOthers {.}}{%
He%
\ \protect \BOthers {.}}{%
{\protect \APACyear {2024}}%
}]{%
He2024}
\APACinsertmetastar {%
He2024}%
\begin{APACrefauthors}%
He, M.%
, Forbes, J\BPBI M.%
, Jacobi, C.%
, Li, G.%
, Liu, L.%
, Stober, G.%
\BCBL {}\ \BBA {} Wang, C.%
\end{APACrefauthors}%
\unskip\
\newblock
\APACrefYearMonthDay{2024}{}{}.
\newblock
{\BBOQ}\APACrefatitle {Observational {Verification} of {High}-{Order} {Solar}
  {Tidal} {Harmonics} in the {Earth}'s {Atmosphere}} {Observational
  {Verification} of {High}-{Order} {Solar} {Tidal} {Harmonics} in the {Earth}'s
  {Atmosphere}}.{\BBCQ}
\newblock
\APACjournalVolNumPages{Geophysical Research Letters}{51}{8}{e2024GL108439}.
\newblock
\begin{APACrefURL}
  [{2024-04-24}]\url{https://onlinelibrary.wiley.com/doi/abs/10.1029/2024GL108439}
  \end{APACrefURL}
\newblock
\APACrefnote{\_eprint:
  https://onlinelibrary.wiley.com/doi/pdf/10.1029/2024GL108439}
\newblock
\begin{APACrefDOI} \doi{10.1029/2024GL108439} \end{APACrefDOI}
\PrintBackRefs{\CurrentBib}

\bibitem [\protect \citeauthoryear {%
He%
, Forbes%
, Li%
, Jacobi%
\BCBL {}\ \BBA {} Hoffmann%
}{%
He%
, Forbes%
\BCBL {}\ \protect \BOthers {.}}{%
{\protect \APACyear {2021}}%
}]{%
He2021g}
\APACinsertmetastar {%
He2021g}%
\begin{APACrefauthors}%
He, M.%
, Forbes, J\BPBI M.%
, Li, G.%
, Jacobi, C.%
\BCBL {}\ \BBA {} Hoffmann, P.%
\end{APACrefauthors}%
\unskip\
\newblock
\APACrefYearMonthDay{2021}{}{}.
\newblock
{\BBOQ}\APACrefatitle {{Mesospheric Q2DW Interactions With Four Migrating Tides
  at 53°N Latitude: Zonal Wavenumber Identification Through Dual-Station
  Approaches}} {{Mesospheric Q2DW Interactions With Four Migrating Tides at
  53°N Latitude: Zonal Wavenumber Identification Through Dual-Station
  Approaches}}.{\BBCQ}
\newblock
\APACjournalVolNumPages{Geophysical Research Letters}{48}{8}{e2020GL092237}.
\newblock
\begin{APACrefURL} \url{https://doi.org/10.1029/2020GL092237} \end{APACrefURL}
\newblock
\begin{APACrefDOI} \doi{10.1029/2020GL092237} \end{APACrefDOI}
\PrintBackRefs{\CurrentBib}

\bibitem [\protect \citeauthoryear {%
He%
, Stober%
\BCBL {}\ \BBA {} Jacobi%
}{%
He%
\ \protect \BOthers {.}}{%
{\protect \APACyear {2023}}%
}]{%
He2023Data}
\APACinsertmetastar {%
He2023Data}%
\begin{APACrefauthors}%
He, M.%
, Stober, G.%
\BCBL {}\ \BBA {} Jacobi, C.%
\end{APACrefauthors}%
\unskip\
\newblock
\APACrefYearMonthDay{2023}{}{}.
\newblock
\APACrefbtitle {{Hourly mesospheric winds over Collm between 2012 and 2020}.}
  {{Hourly mesospheric winds over Collm between 2012 and 2020}.}
\newblock
\APACaddressPublisher{}{Harvard Dataverse}.
\newblock
\begin{APACrefURL}
  \url{{https://dataverse.harvard.edu/privateurl.xhtml?token=ba62bd31-d010-48af-b1d3-85bd8001b1d4}}
  \end{APACrefURL}
\PrintBackRefs{\CurrentBib}

\bibitem [\protect \citeauthoryear {%
He%
, Yamazaki%
\BCBL {}\ \protect \BOthers {.}}{%
He%
, Yamazaki%
\BCBL {}\ \protect \BOthers {.}}{%
{\protect \APACyear {2020}}%
}]{%
He2020jd}
\APACinsertmetastar {%
He2020jd}%
\begin{APACrefauthors}%
He, M.%
, Yamazaki, Y.%
, Hoffmann, P.%
, Hall, C\BPBI M.%
, Tsutsumi, M.%
, Li, G.%
\BCBL {}\ \BBA {} Chau, J\BPBI L.%
\end{APACrefauthors}%
\unskip\
\newblock
\APACrefYearMonthDay{2020}{}{}.
\newblock
{\BBOQ}\APACrefatitle {{Zonal Wave Number Diagnosis of Rossby Wave-Like
  Oscillations Using Paired Ground-Based Radars}} {{Zonal Wave Number Diagnosis
  of Rossby Wave-Like Oscillations Using Paired Ground-Based Radars}}.{\BBCQ}
\newblock
\APACjournalVolNumPages{Journal of Geophysical Research:
  Atmospheres}{125}{12}{}.
\newblock
\begin{APACrefURL} \url{https://doi.org/10.1029/2019JD031599} \end{APACrefURL}
\newblock
\begin{APACrefDOI} \doi{10.1029/2019JD031599} \end{APACrefDOI}
\PrintBackRefs{\CurrentBib}

\bibitem [\protect \citeauthoryear {%
Hocking%
, Fuller%
\BCBL {}\ \BBA {} Vandepeer%
}{%
Hocking%
\ \protect \BOthers {.}}{%
{\protect \APACyear {2001}}%
}]{%
Hocking2001}
\APACinsertmetastar {%
Hocking2001}%
\begin{APACrefauthors}%
Hocking, W\BPBI K.%
, Fuller, B.%
\BCBL {}\ \BBA {} Vandepeer, B.%
\end{APACrefauthors}%
\unskip\
\newblock
\APACrefYearMonthDay{2001}{}{}.
\newblock
{\BBOQ}\APACrefatitle {{Real-time determination of meteor-related parameters
  utilizing modern digital technology}} {{Real-time determination of
  meteor-related parameters utilizing modern digital technology}}.{\BBCQ}
\newblock
\APACjournalVolNumPages{Journal of Atmospheric and Solar-Terrestrial
  Physics}{63}{2}{155--169}.
\newblock
\begin{APACrefURL} \url{https://doi.org/10.1016/S1364-6826(00)00138-3}
  \end{APACrefURL}
\newblock
\begin{APACrefDOI} \doi{10.1016/S1364-6826(00)00138-3} \end{APACrefDOI}
\PrintBackRefs{\CurrentBib}

\bibitem [\protect \citeauthoryear {%
Hocking%
\ \BBA {} Kishore~Kumar%
}{%
Hocking%
\ \BBA {} Kishore~Kumar%
}{%
{\protect \APACyear {2011}}%
}]{%
hocking2011a}
\APACinsertmetastar {%
hocking2011a}%
\begin{APACrefauthors}%
Hocking, W\BPBI K.%
\BCBT {}\ \BBA {} Kishore~Kumar, G.%
\end{APACrefauthors}%
\unskip\
\newblock
\APACrefYearMonthDay{2011}{{\APACmonth{08}}}{}.
\newblock
{\BBOQ}\APACrefatitle {Long term behaviour of the {MLT} quasi-7-day wave at two
  radar-sites at northern polar latitudes} {Long term behaviour of the {MLT}
  quasi-7-day wave at two radar-sites at northern polar latitudes}.{\BBCQ}
\newblock
\APACjournalVolNumPages{Journal of Atmospheric and Solar-Terrestrial
  Physics}{73}{13}{1616--1628}.
\newblock
\begin{APACrefURL}
  [{2024-02-09}]\url{https://www.sciencedirect.com/science/article/pii/S1364682611000393}
  \end{APACrefURL}
\newblock
\begin{APACrefDOI} \doi{10.1016/j.jastp.2011.02.004} \end{APACrefDOI}
\PrintBackRefs{\CurrentBib}

\bibitem [\protect \citeauthoryear {%
Holdsworth%
, Reid%
\BCBL {}\ \BBA {} Cervera%
}{%
Holdsworth%
\ \protect \BOthers {.}}{%
{\protect \APACyear {2004}}%
}]{%
Holdsworth:2004}
\APACinsertmetastar {%
Holdsworth:2004}%
\begin{APACrefauthors}%
Holdsworth, D\BPBI A.%
, Reid, I\BPBI M.%
\BCBL {}\ \BBA {} Cervera, M\BPBI A.%
\end{APACrefauthors}%
\unskip\
\newblock
\APACrefYearMonthDay{2004}{}{}.
\newblock
{\BBOQ}\APACrefatitle {Buckland Park all-sky interferometric meteor radar}
  {Buckland park all-sky interferometric meteor radar}.{\BBCQ}
\newblock
\APACjournalVolNumPages{Radio Science}{39}{5}{n/a--n/a}.
\newblock
\begin{APACrefURL} \url{http://dx.doi.org/10.1029/2003RS003014}
  \end{APACrefURL}
\newblock
\APACrefnote{RS5009}
\newblock
\begin{APACrefDOI} \doi{10.1029/2003RS003014} \end{APACrefDOI}
\PrintBackRefs{\CurrentBib}

\bibitem [\protect \citeauthoryear {%
Holton%
\ \BBA {} Hakim%
}{%
Holton%
\ \BBA {} Hakim%
}{%
{\protect \APACyear {2012}}%
}]{%
Holton2013}
\APACinsertmetastar {%
Holton2013}%
\begin{APACrefauthors}%
Holton, J\BPBI R.%
\BCBT {}\ \BBA {} Hakim, G\BPBI J.%
\end{APACrefauthors}%
\unskip\
\newblock
\APACrefYear{2012}.
\newblock
\APACrefbtitle {{an Introduction To Dynamic Meteorology}} {{an Introduction To
  Dynamic Meteorology}}\ (\BVOL~41)\ (\BNUM~5).
\PrintBackRefs{\CurrentBib}

\bibitem [\protect \citeauthoryear {%
Huang%
\ \protect \BOthers {.}}{%
Huang%
\ \protect \BOthers {.}}{%
{\protect \APACyear {2013}}%
}]{%
huang2013a}
\APACinsertmetastar {%
huang2013a}%
\begin{APACrefauthors}%
Huang, K\BPBI M.%
, Liu, A\BPBI Z.%
, Zhang, S\BPBI D.%
, Yi, F.%
, Huang, C\BPBI M.%
, Gan, Q.%
\BDBL {}Zhang, Y\BPBI H.%
\end{APACrefauthors}%
\unskip\
\newblock
\APACrefYearMonthDay{2013}{{\APACmonth{11}}}{}.
\newblock
{\BBOQ}\APACrefatitle {A nonlinear interaction event between a 16-day wave and
  a diurnal tide from meteor radar observations} {A nonlinear interaction event
  between a 16-day wave and a diurnal tide from meteor radar
  observations}.{\BBCQ}
\newblock
\APACjournalVolNumPages{Annales Geophysicae}{31}{11}{2039--2048}.
\newblock
\begin{APACrefURL}
  [{2024-04-24}]\url{https://angeo.copernicus.org/articles/31/2039/2013/}
  \end{APACrefURL}
\newblock
\begin{APACrefDOI} \doi{10.5194/angeo-31-2039-2013} \end{APACrefDOI}
\PrintBackRefs{\CurrentBib}

\bibitem [\protect \citeauthoryear {%
Jacobi%
}{%
Jacobi%
}{%
{\protect \APACyear {2012}}%
}]{%
Jacobi2012}
\APACinsertmetastar {%
Jacobi2012}%
\begin{APACrefauthors}%
Jacobi, C.%
\end{APACrefauthors}%
\unskip\
\newblock
\APACrefYearMonthDay{2012}{}{}.
\newblock
{\BBOQ}\APACrefatitle {{6 year mean prevailing winds and tides measured by VHF
  meteor radar over Collm (51.3N, 13.0E)}} {{6 year mean prevailing winds and
  tides measured by VHF meteor radar over Collm (51.3N, 13.0E)}}.{\BBCQ}
\newblock
\APACjournalVolNumPages{Journal of Atmospheric and Solar-Terrestrial
  Physics}{78-79}{}{8--18}.
\newblock
\begin{APACrefURL}
  \url{http://www.sciencedirect.com/science/article/pii/S1364682611001210}
  \end{APACrefURL}
\newblock
\begin{APACrefDOI} \doi{https://doi.org/10.1016/j.jastp.2011.04.010}
  \end{APACrefDOI}
\PrintBackRefs{\CurrentBib}

\bibitem [\protect \citeauthoryear {%
Jacobi%
\ \protect \BOthers {.}}{%
Jacobi%
\ \protect \BOthers {.}}{%
{\protect \APACyear {2009}}%
}]{%
jacobi2009b}
\APACinsertmetastar {%
jacobi2009b}%
\begin{APACrefauthors}%
Jacobi, C.%
, Fr{\"o}hlich, K.%
, Portnyagin, Y.%
, Merzlyakov, E.%
, Solovjova, T.%
, Makarov, N.%
\BDBL {}K{\"u}rschner, D.%
\end{APACrefauthors}%
\unskip\
\newblock
\APACrefYearMonthDay{2009}{}{}.
\newblock
{\BBOQ}\APACrefatitle {Semi-empirical model of middle atmosphere wind from the
  ground to the lower thermosphere} {Semi-empirical model of middle atmosphere
  wind from the ground to the lower thermosphere}.{\BBCQ}
\newblock
\APACjournalVolNumPages{Adv. Space Res.}{43}{}{239-246}.
\newblock
\begin{APACrefDOI} \doi{10.1016/j.asr.2008.05.011} \end{APACrefDOI}
\PrintBackRefs{\CurrentBib}

\bibitem [\protect \citeauthoryear {%
Jacobi%
, Hoffmann%
\BCBL {}\ \BBA {} K\"urschner%
}{%
Jacobi%
\ \protect \BOthers {.}}{%
{\protect \APACyear {2008}}%
}]{%
Jacobi2008}
\APACinsertmetastar {%
Jacobi2008}%
\begin{APACrefauthors}%
Jacobi, C.%
, Hoffmann, P.%
\BCBL {}\ \BBA {} K\"urschner, D.%
\end{APACrefauthors}%
\unskip\
\newblock
\APACrefYearMonthDay{2008}{}{}.
\newblock
{\BBOQ}\APACrefatitle {Trends in MLT region winds and planetary waves, Collm
  (52$^{\circ}$N, 15$^{\circ}$E)} {Trends in mlt region winds and planetary
  waves, collm (52$^{\circ}$n, 15$^{\circ}$e)}.{\BBCQ}
\newblock
\APACjournalVolNumPages{Ann. Geophys.}{26}{5}{1221--1232}.
\newblock
\begin{APACrefDOI} \doi{10.5194/angeo-26-1221-2008} \end{APACrefDOI}
\PrintBackRefs{\CurrentBib}

\bibitem [\protect \citeauthoryear {%
Jiang%
\ \protect \BOthers {.}}{%
Jiang%
\ \protect \BOthers {.}}{%
{\protect \APACyear {2008}}%
}]{%
Jiang2008}
\APACinsertmetastar {%
Jiang2008}%
\begin{APACrefauthors}%
Jiang, G.%
, Xu, J.%
, Xiong, J.%
, Ma, R.%
, Ning, B.%
, Murayama, Y.%
\BDBL {}Franke, S\BPBI J.%
\end{APACrefauthors}%
\unskip\
\newblock
\APACrefYearMonthDay{2008}{}{}.
\newblock
{\BBOQ}\APACrefatitle {{A case study of the mesospheric 6.5-day wave observed
  by radar systems}} {{A case study of the mesospheric 6.5-day wave observed by
  radar systems}}.{\BBCQ}
\newblock
\APACjournalVolNumPages{Journal of Geophysical Research
  Atmospheres}{113}{16}{1--12}.
\newblock
\begin{APACrefDOI} \doi{10.1029/2008JD009907} \end{APACrefDOI}
\PrintBackRefs{\CurrentBib}

\bibitem [\protect \citeauthoryear {%
Ma%
\ \protect \BOthers {.}}{%
Ma%
\ \protect \BOthers {.}}{%
{\protect \APACyear {2024}}%
}]{%
Ma2024}
\APACinsertmetastar {%
Ma2024}%
\begin{APACrefauthors}%
Ma, Z.%
, Gong, Y.%
, Zhang, S.%
, Xiao, Q.%
, Huang, C.%
\BCBL {}\ \BBA {} Huang, K.%
\end{APACrefauthors}%
\unskip\
\newblock
\APACrefYearMonthDay{2024}{}{}.
\newblock
{\BBOQ}\APACrefatitle {Quasi-5-{Day} {Oscillations} {During} {Arctic} {Major}
  {Sudden} {Stratospheric} {Warmings} {From} 2005 to 2021} {Quasi-5-{Day}
  {Oscillations} {During} {Arctic} {Major} {Sudden} {Stratospheric} {Warmings}
  {From} 2005 to 2021}.{\BBCQ}
\newblock
\APACjournalVolNumPages{Journal of Geophysical Research: Space
  Physics}{129}{4}{e2023JA032292}.
\newblock
\begin{APACrefURL}
  [{2024-04-26}]\url{https://onlinelibrary.wiley.com/doi/abs/10.1029/2023JA032292}
  \end{APACrefURL}
\newblock
\APACrefnote{\_eprint:
  https://onlinelibrary.wiley.com/doi/pdf/10.1029/2023JA032292}
\newblock
\begin{APACrefDOI} \doi{10.1029/2023JA032292} \end{APACrefDOI}
\PrintBackRefs{\CurrentBib}

\bibitem [\protect \citeauthoryear {%
Matsuno%
}{%
Matsuno%
}{%
{\protect \APACyear {1971}}%
}]{%
Matsuno1971}
\APACinsertmetastar {%
Matsuno1971}%
\begin{APACrefauthors}%
Matsuno, T.%
\end{APACrefauthors}%
\unskip\
\newblock
\APACrefYearMonthDay{1971}{nov}{}.
\newblock
{\BBOQ}\APACrefatitle {{A Dynamical Model of the Stratospheric Sudden Warming}}
  {{A Dynamical Model of the Stratospheric Sudden Warming}}.{\BBCQ}
\newblock
\APACjournalVolNumPages{Journal of the Atmospheric
  Sciences}{28}{8}{1479--1494}.
\newblock
\begin{APACrefURL}
  \url{http://journals.ametsoc.org/doi/abs/10.1175/1520-0469{\%}281971{\%}29028{\%}3C1479{\%}3AADMOTS{\%}3E2.0.CO{\%}3B2}
  \end{APACrefURL}
\newblock
\begin{APACrefDOI} \doi{10.1175/1520-0469(1971)028<1479:ADMOTS>2.0.CO;2}
  \end{APACrefDOI}
\PrintBackRefs{\CurrentBib}

\bibitem [\protect \citeauthoryear {%
McDonald%
, Hibbins%
\BCBL {}\ \BBA {} Jarvis%
}{%
McDonald%
\ \protect \BOthers {.}}{%
{\protect \APACyear {2011}}%
}]{%
McDonald2011}
\APACinsertmetastar {%
McDonald2011}%
\begin{APACrefauthors}%
McDonald, A\BPBI J.%
, Hibbins, R\BPBI E.%
\BCBL {}\ \BBA {} Jarvis, M\BPBI J.%
\end{APACrefauthors}%
\unskip\
\newblock
\APACrefYearMonthDay{2011}{}{}.
\newblock
{\BBOQ}\APACrefatitle {{Properties of the quasi 16 day wave derived from EOS
  MLS observations}} {{Properties of the quasi 16 day wave derived from EOS MLS
  observations}}.{\BBCQ}
\newblock
\APACjournalVolNumPages{Journal of Geophysical Research
  Atmospheres}{116}{6}{1--16}.
\newblock
\begin{APACrefDOI} \doi{10.1029/2010JD014719} \end{APACrefDOI}
\PrintBackRefs{\CurrentBib}

\bibitem [\protect \citeauthoryear {%
Pancheva%
}{%
Pancheva%
}{%
{\protect \APACyear {2001}}%
}]{%
Pancheva2001}
\APACinsertmetastar {%
Pancheva2001}%
\begin{APACrefauthors}%
Pancheva, D.%
\end{APACrefauthors}%
\unskip\
\newblock
\APACrefYearMonthDay{2001}{}{}.
\newblock
{\BBOQ}\APACrefatitle {{Non-linear interaction of tides and planetary waves in
  the mesosphere and lower thermosphere: Observations over Europe}}
  {{Non-linear interaction of tides and planetary waves in the mesosphere and
  lower thermosphere: Observations over Europe}}.{\BBCQ}
\newblock
\APACjournalVolNumPages{Physics and Chemistry of the Earth, Part C: Solar,
  Terrestrial and Planetary Science}{26}{6}{411--418}.
\newblock
\begin{APACrefDOI} \doi{10.1016/S1464-1917(01)00022-8} \end{APACrefDOI}
\PrintBackRefs{\CurrentBib}

\bibitem [\protect \citeauthoryear {%
Pancheva%
\ \BBA {} Mitchell%
}{%
Pancheva%
\ \BBA {} Mitchell%
}{%
{\protect \APACyear {2004}}%
}]{%
Pancheva2004}
\APACinsertmetastar {%
Pancheva2004}%
\begin{APACrefauthors}%
Pancheva, D.%
\BCBT {}\ \BBA {} Mitchell, N\BPBI J.%
\end{APACrefauthors}%
\unskip\
\newblock
\APACrefYearMonthDay{2004}{}{}.
\newblock
{\BBOQ}\APACrefatitle {{Planetary waves and variability of the semidiurnal tide
  in the mesosphere and lower thermosphere over Esrange (68??N, 21??E) during
  winter}} {{Planetary waves and variability of the semidiurnal tide in the
  mesosphere and lower thermosphere over Esrange (68??N, 21??E) during
  winter}}.{\BBCQ}
\newblock
\APACjournalVolNumPages{Journal of Geophysical Research: Space
  Physics}{109}{A8}{}.
\newblock
\begin{APACrefDOI} \doi{10.1029/2004JA010433} \end{APACrefDOI}
\PrintBackRefs{\CurrentBib}

\bibitem [\protect \citeauthoryear {%
Pancheva%
\ \protect \BOthers {.}}{%
Pancheva%
\ \protect \BOthers {.}}{%
{\protect \APACyear {2008}}%
}]{%
Pancheva2008jgr}
\APACinsertmetastar {%
Pancheva2008jgr}%
\begin{APACrefauthors}%
Pancheva, D.%
, Mukhtarov, P.%
, Mitchell, N\BPBI J.%
, Merzlyakov, E.%
, Smith, A\BPBI K.%
, Andonov, B.%
\BDBL {}Murayama, Y.%
\end{APACrefauthors}%
\unskip\
\newblock
\APACrefYearMonthDay{2008}{}{}.
\newblock
{\BBOQ}\APACrefatitle {{Planetary waves in coupling the stratosphere and
  mesosphere during the major stratospheric warming in 2003/2004}} {{Planetary
  waves in coupling the stratosphere and mesosphere during the major
  stratospheric warming in 2003/2004}}.{\BBCQ}
\newblock
\APACjournalVolNumPages{Journal of Geophysical Research
  Atmospheres}{113}{12}{1--22}.
\newblock
\begin{APACrefURL} \url{http://dx.doi.org/10.1029/2007JD009011}
  \end{APACrefURL}
\newblock
\begin{APACrefDOI} \doi{10.1029/2007JD009011} \end{APACrefDOI}
\PrintBackRefs{\CurrentBib}

\bibitem [\protect \citeauthoryear {%
Pogoreltsev%
\ \protect \BOthers {.}}{%
Pogoreltsev%
\ \protect \BOthers {.}}{%
{\protect \APACyear {2002}}%
}]{%
Pogoreltsev2002a}
\APACinsertmetastar {%
Pogoreltsev2002a}%
\begin{APACrefauthors}%
Pogoreltsev, A\BPBI I.%
, Fedulina, I\BPBI N.%
, Mitchell, N\BPBI J.%
, Muller, H\BPBI G.%
, Luo, Y.%
, Meek, C\BPBI E.%
\BCBL {}\ \BBA {} Manson, A\BPBI H.%
\end{APACrefauthors}%
\unskip\
\newblock
\APACrefYearMonthDay{2002}{}{}.
\newblock
{\BBOQ}\APACrefatitle {{Global free oscillations of the atmosphere and
  secondary planetary waves in the mesosphere and lower thermosphere region
  during August/September time conditions}} {{Global free oscillations of the
  atmosphere and secondary planetary waves in the mesosphere and lower
  thermosphere region during August/September time conditions}}.{\BBCQ}
\newblock
\APACjournalVolNumPages{Journal of Geophysical Research
  Atmospheres}{107}{24}{ACL 24--1--ACL 24--12}.
\newblock
\begin{APACrefURL} \url{https://doi.org/10.1029/2001JD001535} \end{APACrefURL}
\newblock
\begin{APACrefDOI} \doi{10.1029/2001JD001535} \end{APACrefDOI}
\PrintBackRefs{\CurrentBib}

\bibitem [\protect \citeauthoryear {%
Riggin%
\ \protect \BOthers {.}}{%
Riggin%
\ \protect \BOthers {.}}{%
{\protect \APACyear {2006}}%
}]{%
Talaat2001}
\APACinsertmetastar {%
Talaat2001}%
\begin{APACrefauthors}%
Riggin, D\BPBI M.%
, Liu, H\BPBI L.%
, Lieberman, R\BPBI S.%
, Roble, R\BPBI G.%
, Russell, J\BPBI M.%
, Mertens, C\BPBI J.%
\BDBL {}Vincent, R\BPBI A.%
\end{APACrefauthors}%
\unskip\
\newblock
\APACrefYearMonthDay{2006}{}{}.
\newblock
{\BBOQ}\APACrefatitle {{Observations of the 5-day wave in the mesosphere and
  lower thermosphere}} {{Observations of the 5-day wave in the mesosphere and
  lower thermosphere}}.{\BBCQ}
\newblock
\APACjournalVolNumPages{Journal of Atmospheric and Solar-Terrestrial
  Physics}{68}{3-5}{323--339}.
\newblock
\begin{APACrefDOI} \doi{10.1016/j.jastp.2005.05.010} \end{APACrefDOI}
\PrintBackRefs{\CurrentBib}

\bibitem [\protect \citeauthoryear {%
Rossby%
}{%
Rossby%
}{%
{\protect \APACyear {1939}}%
}]{%
Rossby1939}
\APACinsertmetastar {%
Rossby1939}%
\begin{APACrefauthors}%
Rossby, C\BHBI G.%
\end{APACrefauthors}%
\unskip\
\newblock
\APACrefYearMonthDay{1939}{}{}.
\newblock
{\BBOQ}\APACrefatitle {{Relation between variations in the intensity of the
  zonal circulation of the atmosphere and the displacements of the
  semi-permanent centers of action}} {{Relation between variations in the
  intensity of the zonal circulation of the atmosphere and the displacements of
  the semi-permanent centers of action}}.{\BBCQ}
\newblock
\APACjournalVolNumPages{Journal of Marine Research}{2}{1}{38--55}.
\newblock
\begin{APACrefDOI} \doi{10.1357/002224039806649023} \end{APACrefDOI}
\PrintBackRefs{\CurrentBib}

\bibitem [\protect \citeauthoryear {%
Salby%
}{%
Salby%
}{%
{\protect \APACyear {1981}}%
}]{%
Salby1981}
\APACinsertmetastar {%
Salby1981}%
\begin{APACrefauthors}%
Salby, M\BPBI L.%
\end{APACrefauthors}%
\unskip\
\newblock
\APACrefYearMonthDay{1981}{}{}.
\newblock
{\BBOQ}\APACrefatitle {{Rossby normal modes in nonuniform background
  configurations. Part II: Equinox and solstice conditions.}} {{Rossby normal
  modes in nonuniform background configurations. Part II: Equinox and solstice
  conditions.}}{\BBCQ}
\newblock
\APACjournalVolNumPages{Journal of the Atmospheric
  Sciences}{38}{9}{1827--1840}.
\newblock
\begin{APACrefDOI} \doi{10.1175/1520-0469(1981)038<1827:RNMINB>2.0.CO;2}
  \end{APACrefDOI}
\PrintBackRefs{\CurrentBib}

\bibitem [\protect \citeauthoryear {%
Sassi%
, Garcia%
\BCBL {}\ \BBA {} Hoppel%
}{%
Sassi%
\ \protect \BOthers {.}}{%
{\protect \APACyear {2012}}%
}]{%
Sassi2012}
\APACinsertmetastar {%
Sassi2012}%
\begin{APACrefauthors}%
Sassi, F.%
, Garcia, R\BPBI R.%
\BCBL {}\ \BBA {} Hoppel, K\BPBI W.%
\end{APACrefauthors}%
\unskip\
\newblock
\APACrefYearMonthDay{2012}{}{}.
\newblock
{\BBOQ}\APACrefatitle {{Large-Scale Rossby Normal Modes during Some Recent
  Northern Hemisphere Winters}} {{Large-Scale Rossby Normal Modes during Some
  Recent Northern Hemisphere Winters}}.{\BBCQ}
\newblock
\APACjournalVolNumPages{Journal of the Atmospheric Sciences}{69}{3}{820--839}.
\newblock
\begin{APACrefDOI} \doi{10.1175/jas-d-11-0103.1} \end{APACrefDOI}
\PrintBackRefs{\CurrentBib}

\bibitem [\protect \citeauthoryear {%
Shi%
\ \protect \BOthers {.}}{%
Shi%
\ \protect \BOthers {.}}{%
{\protect \APACyear {2021}}%
}]{%
shi2021}
\APACinsertmetastar {%
shi2021}%
\begin{APACrefauthors}%
Shi, Y.%
, Evtushevsky, O.%
, Shulga, V.%
, Milinevsky, G.%
, Klekociuk, A.%
, Andrienko, Y.%
\BCBL {}\ \BBA {} Han, W.%
\end{APACrefauthors}%
\unskip\
\newblock
\APACrefYearMonthDay{2021}{{\APACmonth{01}}}{}.
\newblock
{\BBOQ}\APACrefatitle {Mid-{Latitude} {Mesospheric} {Zonal} {Wave} 1 and {Wave}
  2 in {Recent} {Boreal} {Winters}} {Mid-{Latitude} {Mesospheric} {Zonal}
  {Wave} 1 and {Wave} 2 in {Recent} {Boreal} {Winters}}.{\BBCQ}
\newblock
\APACjournalVolNumPages{Remote Sensing}{13}{18}{3749}.
\newblock
\begin{APACrefURL}
  [{2024-05-30}]\url{https://www.mdpi.com/2072-4292/13/18/3749}
  \end{APACrefURL}
\newblock
\APACrefnote{Number: 18 Publisher: Multidisciplinary Digital Publishing
  Institute}
\newblock
\begin{APACrefDOI} \doi{10.3390/rs13183749} \end{APACrefDOI}
\PrintBackRefs{\CurrentBib}

\bibitem [\protect \citeauthoryear {%
Smith%
}{%
Smith%
}{%
{\protect \APACyear {1997}}%
}]{%
smith1997}
\APACinsertmetastar {%
smith1997}%
\begin{APACrefauthors}%
Smith, A\BPBI K.%
\end{APACrefauthors}%
\unskip\
\newblock
\APACrefYearMonthDay{1997}{{\APACmonth{08}}}{}.
\newblock
{\BBOQ}\APACrefatitle {Stationary {Planetary} {Waves} in {Upper} {Mesospheric}
  {Winds}} {Stationary {Planetary} {Waves} in {Upper} {Mesospheric}
  {Winds}}.{\BBCQ}
\newblock
\APACjournalVolNumPages{Journal of the Atmospheric
  Sciences}{54}{16}{2129--2145}.
\newblock
\begin{APACrefURL}
  [{2024-05-13}]\url{https://journals.ametsoc.org/view/journals/atsc/54/16/1520-0469_1997_054_2129_spwium_2.0.co_2.xml}
  \end{APACrefURL}
\newblock
\APACrefnote{Publisher: American Meteorological Society Section: Journal of the
  Atmospheric Sciences}
\newblock
\begin{APACrefDOI} \doi{10.1175/1520-0469(1997)054<2129:SPWIUM>2.0.CO;2}
  \end{APACrefDOI}
\PrintBackRefs{\CurrentBib}

\bibitem [\protect \citeauthoryear {%
Sridharan%
, Tsuda%
, Nakamura%
, Vincent%
\BCBL {}\ \BBA {} {Effendy}%
}{%
Sridharan%
\ \protect \BOthers {.}}{%
{\protect \APACyear {2006}}%
}]{%
sridharan2006}
\APACinsertmetastar {%
sridharan2006}%
\begin{APACrefauthors}%
Sridharan, S.%
, Tsuda, T.%
, Nakamura, T.%
, Vincent, R\BPBI A.%
\BCBL {}\ \BBA {} {Effendy}.%
\end{APACrefauthors}%
\unskip\
\newblock
\APACrefYearMonthDay{2006}{}{}.
\newblock
{\BBOQ}\APACrefatitle {A {Report} on {Radar} {Observations} of 5-8-day {Waves}
  in the {Equatorial} {MLT} {Region}} {A {Report} on {Radar} {Observations} of
  5-8-day {Waves} in the {Equatorial} {MLT} {Region}}.{\BBCQ}
\newblock
\APACjournalVolNumPages{Journal of the Meteorological Society of Japan. Ser.
  II}{84A}{}{295--304}.
\newblock
\begin{APACrefDOI} \doi{10.2151/jmsj.84A.295} \end{APACrefDOI}
\PrintBackRefs{\CurrentBib}

\bibitem [\protect \citeauthoryear {%
Stober%
\ \protect \BOthers {.}}{%
Stober%
\ \protect \BOthers {.}}{%
{\protect \APACyear {2021}}%
}]{%
Stober2021}
\APACinsertmetastar {%
Stober2021}%
\begin{APACrefauthors}%
Stober, G.%
, Kuchar, A.%
, Pokhotelov, D.%
, Liu, H.%
, Liu, H\BHBI L.%
, Schmidt, H.%
\BDBL {}Mitchell, N.%
\end{APACrefauthors}%
\unskip\
\newblock
\APACrefYearMonthDay{2021}{}{}.
\newblock
{\BBOQ}\APACrefatitle {{Interhemispheric differences of mesosphere--lower
  thermosphere winds and tides investigated from three whole-atmosphere models
  and meteor radar observations}} {{Interhemispheric differences of
  mesosphere--lower thermosphere winds and tides investigated from three
  whole-atmosphere models and meteor radar observations}}.{\BBCQ}
\newblock
\APACjournalVolNumPages{Atmospheric Chemistry and
  Physics}{21}{18}{13855--13902}.
\newblock
\begin{APACrefURL} \url{https://acp.copernicus.org/articles/21/13855/2021/}
  \end{APACrefURL}
\newblock
\begin{APACrefDOI} \doi{10.5194/acp-21-13855-2021} \end{APACrefDOI}
\PrintBackRefs{\CurrentBib}

\bibitem [\protect \citeauthoryear {%
Stober%
\ \protect \BOthers {.}}{%
Stober%
\ \protect \BOthers {.}}{%
{\protect \APACyear {2022}}%
}]{%
Stober_2022_3DVAR+DIV}
\APACinsertmetastar {%
Stober_2022_3DVAR+DIV}%
\begin{APACrefauthors}%
Stober, G.%
, Liu, A.%
, Kozlovsky, A.%
, Qiao, Z.%
, Kuchar, A.%
, Jacobi, C.%
\BDBL {}Mitchell, N.%
\end{APACrefauthors}%
\unskip\
\newblock
\APACrefYearMonthDay{2022}{}{}.
\newblock
{\BBOQ}\APACrefatitle {Meteor radar vertical wind observation biases and
  mathematical debiasing strategies including the 3DVAR+DIV algorithm} {Meteor
  radar vertical wind observation biases and mathematical debiasing strategies
  including the 3dvar+div algorithm}.{\BBCQ}
\newblock
\APACjournalVolNumPages{Atmospheric Measurement
  Techniques}{15}{19}{5769--5792}.
\newblock
\begin{APACrefURL} \url{https://amt.copernicus.org/articles/15/5769/2022/}
  \end{APACrefURL}
\newblock
\begin{APACrefDOI} \doi{10.5194/amt-15-5769-2022} \end{APACrefDOI}
\PrintBackRefs{\CurrentBib}

\bibitem [\protect \citeauthoryear {%
Teitelbaum%
\ \BBA {} Vial%
}{%
Teitelbaum%
\ \BBA {} Vial%
}{%
{\protect \APACyear {1991}}%
}]{%
Teitelbaum1991}
\APACinsertmetastar {%
Teitelbaum1991}%
\begin{APACrefauthors}%
Teitelbaum, H.%
\BCBT {}\ \BBA {} Vial, F.%
\end{APACrefauthors}%
\unskip\
\newblock
\APACrefYearMonthDay{1991}{aug}{}.
\newblock
{\BBOQ}\APACrefatitle {{On tidal variability induced by nonlinear interaction
  with planetary waves}} {{On tidal variability induced by nonlinear
  interaction with planetary waves}}.{\BBCQ}
\newblock
\APACjournalVolNumPages{Journal of Geophysical Research: Space
  Physics}{96}{A8}{14169--14178}.
\newblock
\begin{APACrefURL} \url{http://doi.wiley.com/10.1029/91JA01019}
  \end{APACrefURL}
\newblock
\begin{APACrefDOI} \doi{10.1029/91ja01019} \end{APACrefDOI}
\PrintBackRefs{\CurrentBib}

\bibitem [\protect \citeauthoryear {%
van Caspel%
\ \protect \BOthers {.}}{%
van Caspel%
\ \protect \BOthers {.}}{%
{\protect \APACyear {2023}}%
}]{%
Caspel_2023_NAVGEM_tides}
\APACinsertmetastar {%
Caspel_2023_NAVGEM_tides}%
\begin{APACrefauthors}%
van Caspel, W\BPBI E.%
, Espy, P.%
, Hibbins, R.%
, Stober, G.%
, Brown, P.%
, Jacobi, C.%
\BCBL {}\ \BBA {} Kero, J.%
\end{APACrefauthors}%
\unskip\
\newblock
\APACrefYearMonthDay{2023}{}{}.
\newblock
{\BBOQ}\APACrefatitle {A Case Study of the Solar and Lunar Semidiurnal Tide
  Response to the 2013 Sudden Stratospheric Warming} {A case study of the solar
  and lunar semidiurnal tide response to the 2013 sudden stratospheric
  warming}.{\BBCQ}
\newblock
\APACjournalVolNumPages{Journal of Geophysical Research: Space
  Physics}{128}{9}{e2023JA031680}.
\newblock
\begin{APACrefURL}
  \url{https://agupubs.onlinelibrary.wiley.com/doi/abs/10.1029/2023JA031680}
  \end{APACrefURL}
\newblock
\APACrefnote{e2023JA031680 2023JA031680}
\newblock
\begin{APACrefDOI} \doi{https://doi.org/10.1029/2023JA031680} \end{APACrefDOI}
\PrintBackRefs{\CurrentBib}

\bibitem [\protect \citeauthoryear {%
Yamazaki%
\ \BBA {} Matthias%
}{%
Yamazaki%
\ \BBA {} Matthias%
}{%
{\protect \APACyear {2019}}%
}]{%
Yamazaki2019}
\APACinsertmetastar {%
Yamazaki2019}%
\begin{APACrefauthors}%
Yamazaki, Y.%
\BCBT {}\ \BBA {} Matthias, V.%
\end{APACrefauthors}%
\unskip\
\newblock
\APACrefYearMonthDay{2019}{}{}.
\newblock
{\BBOQ}\APACrefatitle {{Large Amplitude Quasi-10-day Waves in the Middle
  Atmosphere during Final Warmings}} {{Large Amplitude Quasi-10-day Waves in
  the Middle Atmosphere during Final Warmings}}.{\BBCQ}
\newblock
\APACjournalVolNumPages{J.Geophys. Res.}{}{}{1--19}.
\newblock
\begin{APACrefDOI} \doi{10.1029/2019JD030634} \end{APACrefDOI}
\PrintBackRefs{\CurrentBib}

\bibitem [\protect \citeauthoryear {%
Yamazaki%
, Matthias%
\BCBL {}\ \BBA {} Miyoshi%
}{%
Yamazaki%
\ \protect \BOthers {.}}{%
{\protect \APACyear {2021}}%
}]{%
yamazaki2021}
\APACinsertmetastar {%
yamazaki2021}%
\begin{APACrefauthors}%
Yamazaki, Y.%
, Matthias, V.%
\BCBL {}\ \BBA {} Miyoshi, Y.%
\end{APACrefauthors}%
\unskip\
\newblock
\APACrefYearMonthDay{2021}{}{}.
\newblock
{\BBOQ}\APACrefatitle {Quasi-4-{Day} {Wave}: {Atmospheric} {Manifestation} of
  the {First} {Symmetric} {Rossby} {Normal} {Mode} of {Zonal} {Wavenumber} 2}
  {Quasi-4-{Day} {Wave}: {Atmospheric} {Manifestation} of the {First}
  {Symmetric} {Rossby} {Normal} {Mode} of {Zonal} {Wavenumber} 2}.{\BBCQ}
\newblock
\APACjournalVolNumPages{Journal of Geophysical Research:
  Atmospheres}{126}{13}{e2021JD034855}.
\newblock
\begin{APACrefDOI} \doi{10.1029/2021JD034855} \end{APACrefDOI}
\PrintBackRefs{\CurrentBib}

\bibitem [\protect \citeauthoryear {%
F\BPBI R.~Yu%
\ \protect \BOthers {.}}{%
F\BPBI R.~Yu%
\ \protect \BOthers {.}}{%
{\protect \APACyear {2019}}%
}]{%
yu2019a}
\APACinsertmetastar {%
yu2019a}%
\begin{APACrefauthors}%
Yu, F\BPBI R.%
, Huang, K\BPBI M.%
, Zhang, S\BPBI D.%
, Huang, C\BPBI M.%
, Yi, F.%
, Gong, Y.%
\BDBL {}Ning, B.%
\end{APACrefauthors}%
\unskip\
\newblock
\APACrefYearMonthDay{2019}{}{}.
\newblock
{\BBOQ}\APACrefatitle {Quasi 10- and 16-{Day} {Wave} {Activities} {Observed}
  {Through} {Meteor} {Radar} and {MST} {Radar} {During} {Stratospheric} {Final}
  {Warming} in 2015 {Spring}} {Quasi 10- and 16-{Day} {Wave} {Activities}
  {Observed} {Through} {Meteor} {Radar} and {MST} {Radar} {During}
  {Stratospheric} {Final} {Warming} in 2015 {Spring}}.{\BBCQ}
\newblock
\APACjournalVolNumPages{Journal of Geophysical Research:
  Atmospheres}{124}{12}{6040--6056}.
\newblock
\begin{APACrefURL}
  [{2024-02-09}]\url{https://onlinelibrary.wiley.com/doi/abs/10.1029/2019JD030630}
  \end{APACrefURL}
\newblock
\APACrefnote{\_eprint:
  https://onlinelibrary.wiley.com/doi/pdf/10.1029/2019JD030630}
\newblock
\begin{APACrefDOI} \doi{10.1029/2019JD030630} \end{APACrefDOI}
\PrintBackRefs{\CurrentBib}

\bibitem [\protect \citeauthoryear {%
Y.~Yu%
\ \protect \BOthers {.}}{%
Y.~Yu%
\ \protect \BOthers {.}}{%
{\protect \APACyear {2013}}%
}]{%
Yu2013}
\APACinsertmetastar {%
Yu2013}%
\begin{APACrefauthors}%
Yu, Y.%
, Wan, W.%
, Ning, B.%
, Liu, L.%
, Wang, Z.%
, Hu, L.%
\BCBL {}\ \BBA {} Ren, Z.%
\end{APACrefauthors}%
\unskip\
\newblock
\APACrefYearMonthDay{2013}{}{}.
\newblock
{\BBOQ}\APACrefatitle {{Tidal wind mapping from observations of a meteor radar
  chain in December 2011}} {{Tidal wind mapping from observations of a meteor
  radar chain in December 2011}}.{\BBCQ}
\newblock
\APACjournalVolNumPages{Journal of Geophysical Research: Space
  Physics}{118}{5}{2321--2332}.
\newblock
\begin{APACrefURL}
  \url{https://agupubs.onlinelibrary.wiley.com/doi/abs/10.1029/2012JA017976}
  \end{APACrefURL}
\newblock
\begin{APACrefDOI} \doi{10.1029/2012JA017976} \end{APACrefDOI}
\PrintBackRefs{\CurrentBib}

\bibitem [\protect \citeauthoryear {%
Zaqarashvili%
\ \protect \BOthers {.}}{%
Zaqarashvili%
\ \protect \BOthers {.}}{%
{\protect \APACyear {2021}}%
}]{%
Zaqarashvili2021}
\APACinsertmetastar {%
Zaqarashvili2021}%
\begin{APACrefauthors}%
Zaqarashvili, T\BPBI V.%
, Albekioni, M.%
, Ballester, J\BPBI L.%
, Bekki, Y.%
, Biancofiore, L.%
, Birch, A\BPBI C.%
\BDBL {}Yellin-Bergovoy, R.%
\end{APACrefauthors}%
\unskip\
\newblock
\APACrefYearMonthDay{2021}{}{}.
\newblock
{\BBOQ}\APACrefatitle {{Rossby Waves in Astrophysics}} {{Rossby Waves in
  Astrophysics}}.{\BBCQ}
\newblock
\APACjournalVolNumPages{Space Science Reviews}{217}{1}{15}.
\newblock
\begin{APACrefURL} \url{https://doi.org/10.1007/s11214-021-00790-2}
  \end{APACrefURL}
\newblock
\begin{APACrefDOI} \doi{10.1007/s11214-021-00790-2} \end{APACrefDOI}
\PrintBackRefs{\CurrentBib}

\bibitem [\protect \citeauthoryear {%
Zhang%
\ \BBA {} Forbes%
}{%
Zhang%
\ \BBA {} Forbes%
}{%
{\protect \APACyear {2014}}%
}]{%
Zhang2014a}
\APACinsertmetastar {%
Zhang2014a}%
\begin{APACrefauthors}%
Zhang, X.%
\BCBT {}\ \BBA {} Forbes, J\BPBI M.%
\end{APACrefauthors}%
\unskip\
\newblock
\APACrefYearMonthDay{2014}{dec}{}.
\newblock
{\BBOQ}\APACrefatitle {{Lunar tide in the thermosphere and weakening of the
  northern polar vortex}} {{Lunar tide in the thermosphere and weakening of the
  northern polar vortex}}.{\BBCQ}
\newblock
\APACjournalVolNumPages{Geophysical Research Letters}{41}{23}{8201--8207}.
\newblock
\begin{APACrefURL} \url{http://doi.wiley.com/10.1002/2014GL062103}
  \end{APACrefURL}
\newblock
\begin{APACrefDOI} \doi{10.1002/2014GL062103} \end{APACrefDOI}
\PrintBackRefs{\CurrentBib}

\bibitem [\protect \citeauthoryear {%
Zhao%
\ \protect \BOthers {.}}{%
Zhao%
\ \protect \BOthers {.}}{%
{\protect \APACyear {2019}}%
}]{%
Zhao2019}
\APACinsertmetastar {%
Zhao2019}%
\begin{APACrefauthors}%
Zhao, Y.%
, Taylor, M\BPBI J.%
, Pautet, P\BHBI D.%
, Moffat-Griffin, T.%
, Hervig, M\BPBI E.%
, Murphy, D\BPBI J.%
\BDBL {}{Russell III}, J\BPBI M.%
\end{APACrefauthors}%
\unskip\
\newblock
\APACrefYearMonthDay{2019}{}{}.
\newblock
{\BBOQ}\APACrefatitle {{Investigating an Unusually Large 28-Day Oscillation in
  Mesospheric Temperature Over Antarctica Using Ground-Based and Satellite
  Measurements}} {{Investigating an Unusually Large 28-Day Oscillation in
  Mesospheric Temperature Over Antarctica Using Ground-Based and Satellite
  Measurements}}.{\BBCQ}
\newblock
\APACjournalVolNumPages{Journal of Geophysical Research:
  Atmospheres}{124}{15}{8576--8593}.
\newblock
\begin{APACrefURL}
  \url{https://agupubs.onlinelibrary.wiley.com/doi/abs/10.1029/2019JD030286}
  \end{APACrefURL}
\newblock
\begin{APACrefDOI} \doi{10.1029/2019JD030286} \end{APACrefDOI}
\PrintBackRefs{\CurrentBib}

\end{thebibliography}

\clearpage

\clearpage
\renewcommand{\arraystretch}{0.5}

\clearpage

\clearpage

\begin{sidewaystable}
\centering
\caption { Novel Findings versus Established Phenomena: Ground-Observed Nonlinear Interactions Among  Atmospheric Planetary-Scale Waves} \label{tab:title}
\centering
\begin{tabular}{c|c|c|c|c|c}
\hline
\rowcolor{gray!20}
Index & \begin{tabular}[c]{@{}c@{}}PDT Results\\  $[\frac{f}{1 \text{cpd}}, s]$\end{tabular}  & \begin{tabular}[c]{@{}c@{}}Explainable\\ Alias \end{tabular}   & \begin{tabular}[c]{@{}c@{}} Potential Interactions:\\$[\frac{f_1}{1 \text{cpd}}, s_1]\pm [\frac{f_2}{1 \text{cpd}}, s_2]=[\frac{f_3}{1 \text{cpd}}, s_3]$\end{tabular} & \begin{tabular}[c]{@{}c@{}}$f$ Matching \\ Evidence \end{tabular} & \begin{tabular}[c]{@{}c@{}} Evidence for\\   $f$ and $s$ Matching\end{tabular} \\ \hline
(a$+$)& $[\frac{1}{6},2]$ & &  $[\frac{1}{6},1]+[0,1]=[\frac{1}{6},2]$ & & \\
(a$-$)& &  $[\frac{1}{6},-1]$ & $[\frac{1}{6},1]-[0,2]=[\frac{1}{6},-1]$ & \multirow{-2}{*}{Meaningless} & \multirow{-2}{*}{First-time disclosure} \\	\hline
\rowcolor{gray!10}
(b$+$)& $[\frac{1}{10},2]$ & &  $[\frac{1}{10},1]+[0,1]=[\frac{1}{10},2]$ & &  \\
\rowcolor{gray!10}
(b$-$)&  &  $[\frac{1}{10},-1]$ & $[\frac{1}{10},1]-[0,2]=[\frac{1}{10},-1]$ & \multirow{-2}{*}{Meaningless} & \multirow{-2}{*}{First-time disclosure } \\  \hline
(c$+$)& $[\frac{1}{16},2]$& &  $[\frac{1}{16},1]+[0,1]=[\frac{1}{16},2]$ & &\citeA{He2020} \\
(c$-$) & & $[\frac{1}{16},-1]$& $[\frac{1}{16},1]-[0,2]=[\frac{1}{16},-1]$ & \multirow{-2}{*}{Meaningless} & First-time proposed \\ 	\hline
\rowcolor{gray!10}
(d)&$[\frac{1}{8},2]$ & & $[\frac{1}{16},1]+[\frac{1}{16},1]=[\frac{1}{8},2]$ & \citeA{He2022NC} & \citeA{He2022NC} \\ 	\hline
(e) &$[1,2]$ & & $[1,1]+[0,1]=[1,2]$ & Meaningless & First-time disclosure \\ 	\hline
\rowcolor{gray!10}
(f) &$[3,4]$ & & $[3,3]+[0,1]=[3,4]$ & Meaningless & First-time disclosure \\ 	\hline
(g)& $[3,2]$& & $[3,3]-[0,1]=[3,2]$ & Meaningless & First-time disclosure \\ 	\hline
\rowcolor{gray!10}
(h) & $[\frac{24}{22.6},2]$& & $[1,1]+[\frac{1}{16},1]=[\frac{24}{22.6},2]$ & \citeA{huang2013a} & First-time disclosure \\ 	\hline
(i)&$[\frac{24}{21.8},2]$ & & $[1,1]+[\frac{1}{10},1]=[\frac{24}{21.8},2]$ &   \multicolumn{1}{c|}{\cellcolor[HTML]{FFFFFF}}   & First-time disclosure \\	\hhline{----~-}
\rowcolor{gray!10}
(j) &$ [\frac{24}{20.6},2]$& & $[1,1]+[\frac{1}{6},1]=[\frac{24}{20.6},2]$ &  \multicolumn{1}{c|}{\cellcolor[HTML]{FFFFFF}}   & First-time disclosure \\ \hhline{----~-}
(k)& $[\frac{24}{20.6},3]$& & $[1,2]+[\frac{1}{6},1]=[\frac{24}{20.6},3]$ &\multicolumn{1}{c|}{\multirow{-3}{*}{\cellcolor[HTML]{FFFFFF}  \citeA{Teitelbaum1991}$^*$}}& First-time disclosure \\	\hline
\rowcolor{gray!10}
(l)& $[\frac{24}{12.4},1]$& & $[2,2]-[\frac{1}{16},1]=[\frac{24}{12.4},1]$ & \citeA{Pancheva2004} & \citeA{He2018g} \\	\hline
(m)& $[\frac{24}{12.6},1]$& & $[2,2]-[\frac{1}{10},1]=[\frac{24}{12.6},1]$ & \citeA{Pancheva2004} & \citeA{He2020GRLb} \\	\hline
\rowcolor{gray!10}
(n) & $[\frac{24}{11.6},3]$ & & $[2,2]+[\frac{1}{16},1]=[\frac{24}{11.6},3]$ & \citeA{Pancheva2001} & \citeA{He2018j} \\	\hline
(o) & $[\frac{24}{11.4},3]$& & $[2,2]+[\frac{1}{10},1]=[\frac{24}{11.4},3]$ & \citeA{Pancheva2001} & \citeA{He2020GRLb} \\	\hline
\rowcolor{gray!10}
(p) & $[\frac{24}{11.1},3]$& & $[2,2]+[\frac{1}{6},1]=[\frac{24}{11.1},3]$ & \citeA{Pancheva2001} & First-time disclosure \\	\hline
(q) &$[\frac{24}{8.2},2]$  & & $[3,3]-[\frac{1}{16},1]=[\frac{24}{8.2},2]$ & First-time disclosure & First-time disclosure \\	\hline
\rowcolor{gray!10}
(r) & $[\frac{24}{8.3},2]$& & $[3,3]-[\frac{1}{10},1]=[\frac{24}{8.3},2]$ & First-time disclosure & First-time disclosure \\	\hline
(s) &$[\frac{24}{8.5},2]$ & & $[3,3]-[\frac{1}{6},1]=[\frac{24}{8.5},2]$ & First-time disclosure & First-time disclosure \\	\hline
\rowcolor{gray!10}
(t) & $[\frac{24}{7.6},4]$& & $[3,3]+[\frac{1}{6},1]=[\frac{24}{7.6},4]$ & First-time disclosure & First-time disclosure \\	\hline
\end{tabular}

\footnotesize{ *The spectral analysis in \citeA{Teitelbaum1991}  did not discern between these interactions  (i), (j) and (k), even only in the frequency domain.  By integrating zonal wavenumber diagnosis and ensuring proper frequency resolution, our current work distinguishes between these interactions.}\\
\end{sidewaystable}
\end{document}